\documentclass[aps,reprint,groupedaddress]{revtex4-1}
\usepackage{amsmath}
\usepackage{amssymb}

\usepackage{ulem}

\usepackage{dsfont}
\usepackage{graphicx}
\usepackage{hyperref}
\usepackage{color}
\newcommand{\FT}[1]{\tilde{#1}}
\newcommand{\tg}{\tau_{\Gamma}}
\newcommand{\td}{\tau_{D}}
\newcommand{\tm}{\tau_{m}}
\newcommand{\tf}{\tau_{\mathrm{MFP}}}

\newcommand{\noise}{\eta}
\newcommand{\kT}{k_{\mathrm{B}}T}

\newcommand{\xDL}{\tilde{x}}
\newcommand{\dotxDL}{\dot{\tilde{x}}}
\newcommand{\ddotxDL}{\ddot{\tilde{x}}}
\newcommand{\tDL}{\tilde{t}}
\newcommand{\FDL}{\tilde{F}}
\newcommand{\noiseDL}{\tilde{\eta}}
\newcommand{\yDL}{\tilde{y}}
\newcommand{\dotyDL}{\dot{\tilde{y}}}

\newcommand{\xiDL}{\tilde{\xi}}

\newcommand{\tgOne}{\tau_{1}}
\newcommand{\tgTwo}{\tau_{2}}
\newcommand{\tgi}{\tau_{i}}

\newcommand{\tgj}{\tau_{j}}
\newcommand{\tgk}{\tau_{k}}
\newcommand{\tgl}{\tau_{l}}

\newcommand{\tfp}{\tau_{\mathrm{FP}}}

\newcommand{\tfovi}{\tau_{\mathrm{OD}}^{(i)}}
\newcommand{\tfedi}{\tau_{\mathrm{ED}}^{(i)}}

\begin{document}

%\preprint{}


%\title{Barrier crossing in the presence of bi-exponential memory is dominated by fast
%orthogonal degrees of freedom}

\title{Non-Markovian barrier crossing with two-time-scale memory is dominated by 
the faster  memory component}


\author{Julian Kappler}
\author{Victor B.~Hinrichsen}
\author{Roland R.~Netz}
%\thanks{}
\affiliation{Freie Universit\"at Berlin}


\date{\today}
\begin{abstract}
% maximally 500 words
We investigate non-Markovian barrier-crossing kinetics of a massive particle in one dimension
 in the presence of a memory function  that is the sum of two exponentials with different
 memory times $\tgOne$ and $\tgTwo$.
\textcolor{black}{
Our Langevin simulations for the special case where both exponentials contribute equally to the 
total friction show that the barrier crossing time becomes independent of the longer memory 
time if at least one of the two memory times is larger than the intrinsic diffusion time.}
When we associate memory effects with coupled degrees of freedom that are orthogonal to a
one-dimensional reaction coordinate,
this counterintuitive result  shows that the  faster orthogonal degrees of freedom  dominate   
 barrier-crossing kinetics  in the non-Markovian limit and that the slower orthogonal  degrees become negligible,
quite contrary to the standard  time-scale separation assumption and with important
consequences for the proper setup  of coarse-graining procedures in the non-Markovian case.
By asymptotic matching and symmetry arguments, we construct a  crossover
formula for the barrier crossing time that is valid for general  multi-exponential memory kernels.
\textcolor{black}{This formula can be used to estimate barrier-crossing times for general memory functions for high
friction, i.e.~in the overdamped regime, as well as for low friction, i.e.~in the inertial regime.}
Typical examples where our results are important
include protein folding in the high-friction limit
and chemical reactions such as proton-transfer reactions in the low-friction limit.
\end{abstract}

%
%
%% insert suggested PACS numbers in braces on next line
%\pacs{}
%% insert suggested keywords - APS authors don't need to do this
%%\keywords{}
%
%%\maketitle must follow title, authors, abstract, \pacs, and \keywords
\maketitle
%
%% body of paper here - Use proper section commands
%% References should be done using the \cite, \ref, and \label commands
%%\section{}
%% Put \label in argument of \section for cross-referencing
%%\section{\label{}}
%%\subsection{}
%%\subsubsection{}
%
%
%\begin{widetext}
%\tableofcontents
%
%
%
%\begin{widetext}
%

\section{Introduction}

Rare events,  such as chemical reactions, macromolecular conformational transitions, or nucleation events,
can be modeled as barrier crossing in an effective  one-dimensional energy
landscape \cite{kramers_brownian_1940,chandler1986,berne1988,hanggi_reaction-rate_1990,
best_diffusive_2006}.
Due to the coupling to intra- and intermolecular orthogonal degrees of freedom,
the dynamics of the reaction coordinate 
becomes  non-Markovian and can be   characterized by a memory function that 
describes for how long a system remembers its past state \cite{zwanzig_memory_1961,mori_transport_1965,Schmid_2017,Schmid_2018,Schilling_2017,Schilling_2019}.
Memory  effects have been studied in the context of ion-pair kinetics \cite{rey_dynamical_1992,mullen_transmission_2014},
conformational transitions in small molecules \cite{rosenberg1980,de_sancho_molecular_2014,daldrop_butane_2018}, 
 diffusive motion of particles and molecules in liquids  \cite{Mason1995,lesnicki_molecular_2016,daldrop_external_2017,Bechinger2018}, 
  cell locomotion  \cite{selmeczi2005},
 polymer looping kinetics \cite{wilemski_diffusioncontrolled_1974,szabo_first_1980,Dua_2011,gowdy_nonexponential_2017,guerin_non-markovian_2012}
 and  protein folding \cite{plotkin_non-markovian_1998,Makarov2018}, and have been demonstrated to
substantially influence 
 barrier-crossing times for slowly decaying  memory functions  \cite{grote_stable_1980,carmeli_non-markoffian_1982,straub_non-markovian_1986,talkner_transition_1988,pollak_theory_1989,ianconescu_study_2015}. 

In previous \textcolor{black}{numerical} studies concerned with \textcolor{black}{position-independent} 
memory effects
  on the barrier-crossing kinetics, 
the memory function has been  assumed to be single exponential or single Gaussian
and is thus chraracterized by a single
time scale, the memory time $\tg$  \cite{straub_non-markovian_1986,pollak_theory_1989,tucker_comparison_1991,
ianconescu_study_2015,kappler_memory-induced_2018}.
Even in this simple scenario the effect  memory has on the  time needed to cross the barrier,
the mean-first passage time $\tf$, is subtle: 
for  a memory time $\tg$  that is longer than the intrinsic diffusion time scale
and for fixed friction coefficient $\gamma$ (defined by the integral over the memory function),
  $\tf$  scales as  $\tf \sim \tg^2 {\rm e}^{\beta U_0}$,
where $U_0$ denotes the barrier height and $\beta=1/(\kT)$ the inverse thermal energy 
 \cite{straub_non-markovian_1986,kappler_memory-induced_2018}.
The exponential term corresponds to the Arrhenius law and is for many applications the  dominant factor. 
In the present context we are mostly interested in  the pre-exponential factor, which signals
 that the presence of  memory modifies the barrier-crossing kinetics even
when the barrier-crossing time  $\tf$  is much longer than the memory time $\tg$;
thus, a naive time-scale separation argument, according to which memory would only influence
reaction kinetics up to times scales of the memory time itself, is not valid.
For intermediate values of $\tg$ a distinct scaling regime exists,  where memory in fact speeds up 
barrier-crossing kinetics, meaning that   $\tf$ is significantly shorter than in the Markovian
limit when $\tg$  tends to zero at fixed $\gamma$  \cite{kappler_memory-induced_2018}.
Thus, i) whether memory speeds up or slows down reaction kinetics depends on the precise value
of the memory time, and ii) for long memory time $\tg$   the barrier-crossing time $\tf$
is influenced by memory effects even when  $\tf \gg \tg$.

While these results vividly demonstrate  the complex influence    of  memory on  reaction kinetics, they
are of only limited practical use, since memory effects are typically caused 
by the reaction coordinate coupling to several 
orthogonal degrees of freedom
and as a consequence memory functions are characterized by more than one time scale
\cite{rey_dynamical_1992,tolokh_prediction_2002,de_sancho_molecular_2014,lesnicki_molecular_2016,gottwald_parametrizing_2015,daldrop_external_2017}.
\textcolor{black}{Systematic numerical} results on the effect of  memory functions with more than
 one time scale on reaction  kinetics  have
not been presented in literature. 

In this  work, we study the barrier-crossing dynamics of a massive particle  in a 
one-dimensional double-well potential 
in the presence of a memory function that consists of the sum of two exponentials
with different decay times $\tgOne$ and $\tgTwo$.
To limit the number of parameters 
in our simulations of the   generalized Langevin equation (GLE),
we consider the case where the two exponentials 
each contribute equally  to the friction coefficient,
i.e.~when the integrals over each memory component are equal. 
Contrary to what one might intuitively expect, 
we find that in the asymmetric case $\tgOne \ll \tgTwo$,
 the barrier-crossing time  is 
\textcolor{black}{ solely determined by
the memory contribution with the shorter memory time $\tgOne$
if  $ \tgTwo$ exceeds
 the intrinsic diffusion time scale;}
this result is also predicted analytically by a pole analysis of the linearized  Langevin
equation. 
Our finding  is  at odds with standard time scale separation arguments
and coarse graining procedures
that are normally designed to project onto the slowest degrees of freedom
 \cite{northrup_stable_1980,zwanzig_nonlinear_1973, Marrink2007, Papoian2013}.
We conclude that for  non-Markovian barrier-crossing phenomena,  in particular 
the coupling to the fast -- not the slow -- orthogonal degrees of freedom 
needs to be preserved in order to obtain an accurate description of the reaction kinetics.

Based on symmetry arguments and asymptotically valid rate expressions, we construct 
a heuristic crossover  formula for the barrier-crossing time  in a double well potential
 in the presence of a  general 
multi-exponential memory kernel.
This crossover  formula not only  accurately  describes our numerical results for 
bi-exponential memory functions, it  can also be  used to estimate the
effects general memory functions have on the reaction kinetics of experimental and simulation
systems. 

{
Our  results  are   relevant for the description of the  reaction kinetics  of general non-Markovian systems. 
For example, pronounced memory effects are expected 
when the  folding of a protein is  described in terms of a reaction coordinate
that does not properly describe all barriers relevant for the folding process  \cite{plotkin_non-markovian_1998}.
\textcolor{black}{The traditional approach would be to construct an improved reaction coordinate
that shows minimal memory effects.
We here consider an alternative  modeling approach 
that includes non-Markovian effects,
and in particular the fast-decaying memory contributions 
in order to correctly describe the reaction kinetics. }
One {other} example where strong memory effects are expected
are fast chemical reactions in a solvent, for example proton-transfer reactions in water, 
where the motion along the reaction coordinate and solvent motion occur on the same time scale
\cite{marx1999,daldrop2018}.
Also here, a reduced description
in terms of a low-dimensional reaction coordinate  becomes valid if  memory effects are properly included. 
Again,  the quickly decaying memory contributions will be particularly important. 
Our paper is a first step towards the \textcolor{black}{systematic} usage of multi-scale memory functions for
 non-Markovian kinetic modeling. }

\section{Model for barrier crossing  with two-time-scale memory}

\begin{figure}[ht]
%\centering
	\includegraphics[width=0.8\columnwidth]{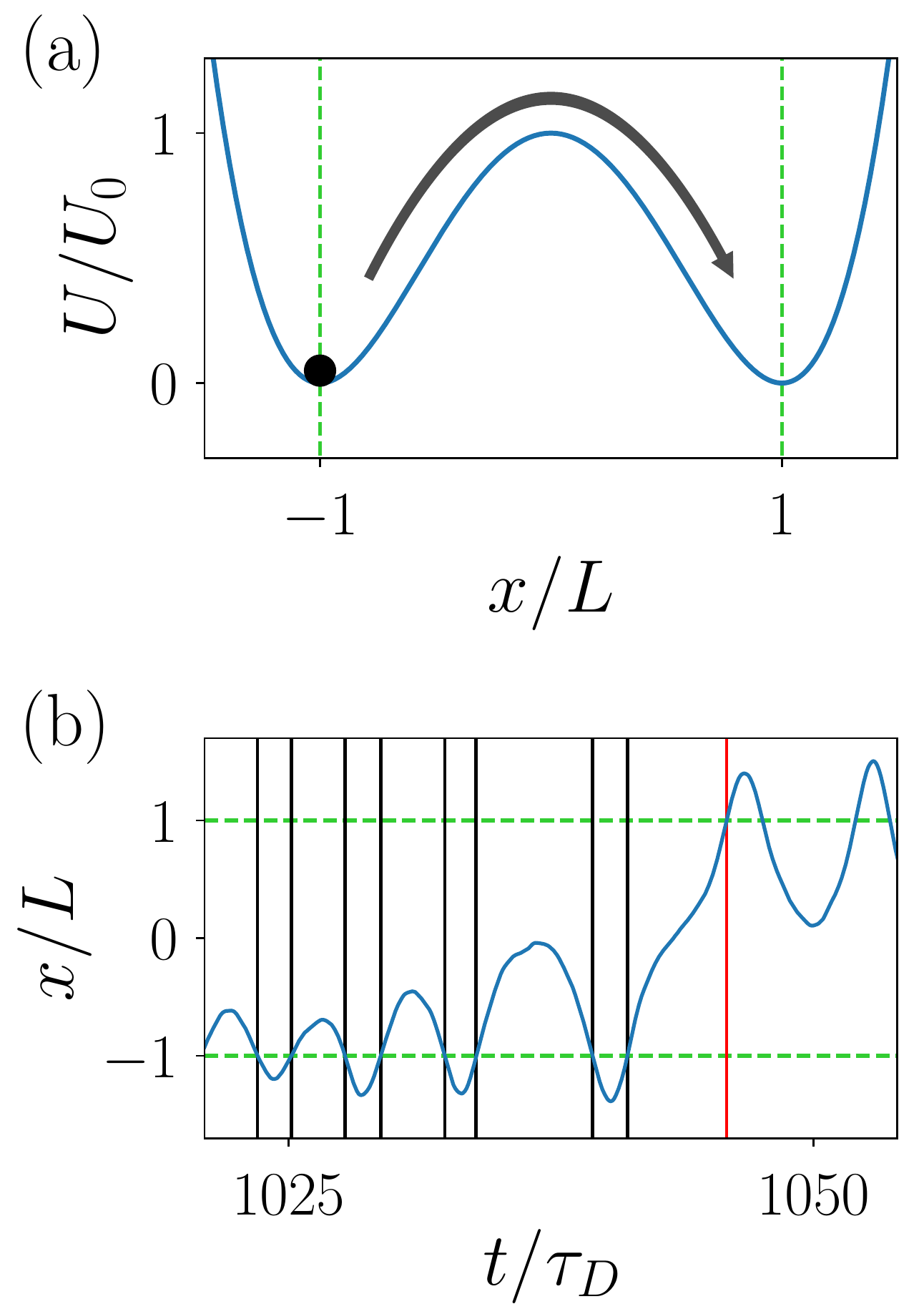}
\caption{
\label{fig:DoubleExp_Potential}
 {(a) Illustration of the barrier crossing} of a massive particle in the double-well potential $U(x)$ given by Eq.~\eqref{eq:DoubleExp_QuarticPotential}.
The mean first-passage time  $\tf$ is defined as the mean time difference 
between crossing the minimum at $x =-L$ (left dashed line) 
and \textcolor{black}{crossing} the other minimum at $x=L$ (right dashed line) for the first time.
{(b) Illustration of how first-passage times (FPTs) are obtained from Langevin simulations.}
Vertical black lines mark crossings of the trajectory with the potential minimum $x = -L$ (lower  dashed horizontal  line), 
the vertical red line marks the first crossing of the trajectory with the potential minimum $x = L$ (upper  dashed horizontal  line).
Each vertical black line contributes to the ensemble of first-passage events
and the corresponding FPT is obtained by calculating 
 the time difference between the black and  the red lines.
The MFPT $\tf$ is obtained by averaging over all FPTs.
For the simulation trajectory shown, 
the parameters $\tm/\td = 10$, $\tgOne/\td = \tgTwo/\td = 0.001$ and barrier height $\beta U_0 = 3$ are used.
}
\end{figure}

We consider the one-dimensional generalized Langevin equation (GLE)
\cite{zwanzig_memory_1961,mori_transport_1965,grote_stable_1980,pollak_theory_1989}
\begin{equation}
	\label{eq:DoubleExp_LangevinDoubleExp}
	m \,\ddot{x}(t) = - \int_0^{t} \Gamma(t-t') \dot{x}(t')\,\mathrm{d}t' -  U'(x(t)) + \noise(t),
\end{equation}
where $m$ is the effective mass of the reaction coordinate $x$, 
$\Gamma(t)$ is the  memory function,
 $U'(x)$ denotes the derivative of the   potential $U(x)$,
 and $\noise(t)$ is a time-dependent random force.
The random force is  Gaussian with zero mean, $\langle \noise(t) \rangle = 0$,
and obeys the generalized fluctuation-dissipation theorem (FDT)
	$\beta \left\langle \noise(t)\noise(t') \right\rangle =  \Gamma(|t-t'|)$.
We consider a symmetric double-well potential
\begin{equation}
	\label{eq:DoubleExp_QuarticPotential}
	U(x) = U_0 \left[ \left(\frac{x}{L}\right)^2 - 1\right]^2,
\end{equation}
illustrated in Fig.~\ref{fig:DoubleExp_Potential} (a),
which is characterized by the barrier height $U_0$ and by the spatial separation 
$2L$ between the two potential wells.
 In our simulations we assume
a bi-exponential memory function 
\begin{align}
	\label{eq:DoubleExp_DoubleExpMemKern}
	\Gamma(t) &= \sum_{i=1}^2 \frac{\gamma_i}{\tgi} \exp\left(-{|t|}/{\tgi}\right) %\\
\end{align}
where $\tgOne$, $\tgTwo$ are two memory time scales and the  corresponding friction coefficients are
$\gamma_1$, $\gamma_2$. We call
$\gamma = \int_0^{\infty} \mathrm{d}t\, \Gamma(t) = \gamma_1 + \gamma_2$ 
 the total friction coefficient.
In our simulations we consider the special case where each of the two exponentials contributes 
 equally to the total friction coefficient, i.e. $\gamma_1 = \gamma_2 = \gamma/2$. %,
To reduce the number of parameters,
we introduce the time scales
\begin{equation}
	\label{eq:DoubleExp_DefTmTd}
	\tm = \frac{m}{\gamma}, \qquad \qquad \td = \beta L^2 \gamma,
\end{equation}
where the inertial
time  $\tm$ characterizes viscous dissipation of particle momentum, 
and $\td$  is the intrinsic diffusion time which depends on  the barrier 
separation $L$ and the  total friction coefficient $\gamma$. 
Memory effects are  important if the memory time exceeds $\td$.
With these definitions, the bi-exponential system with $\gamma_1 = \gamma_2 = \gamma/2$ 
 is determined by the three
 dimensionless time scale ratios $\tm/\td$, $\tgOne/\td$, $\tgTwo/\td$
 and the dimensionless barrier height $\beta U_0$,
 see Appendix \ref{app:DimensionlessGLE} for details.
In our simulations we use a fixed 
barrier height of $\beta U_0 = 3$\textcolor{black}{, except in Appendix \ref{app:BarrierHeight} 
where we also present
results for higher values of $\beta U_0$.}

To simulate the GLE numerically, we explicitly couple the particle coordinate $x$ to   two  auxiliary degrees 
of freedom with  relaxation times $\tgOne$ and $\tgTwo$, see Appendix \ref{app:TransformingToMarkovian} for details.
Using a fourth-order Runge-Kutta integration scheme, we then simulate the composite  system
in the parameter range $\tm/\td \in [10^{-3},10^{3}]$, $ \tgOne/\td,  \tgTwo/\td \in [10^{-3},10^{2}]$.
Initial particle  positions are sampled from a Gaussian approximation of the probability distribution in the 
left well around $x = -L$, i.e., $\langle x(0) \rangle = -L$, $\beta \langle (x(0) + L)^2 \rangle = 1/U''(-L) = L^2 /(8U_0)$.
Initial velocities are sampled from a Gaussian distribution with zero mean and variance 
$\beta \langle \dot{x}^2(0) \rangle = 1/m$, in accordance with the equipartition theorem.
The initial values for the auxiliary  degrees of freedom are sampled from their corresponding equilibrium distributions,
see Appendix \ref{app:TransformingToMarkovian} for details.

From our simulations we obtain  distributions for the first-passage time (FPT) $\tfp$ defined by
the time  difference between 
crossing the potential minimum at $x=-L$ and reaching the potential 
minimum at $x=L$ for the first time, see Fig.~\ref{fig:DoubleExp_Potential} (b)
for an illustration. 
Since the potential is symmetric, we also collect first-passage events from 
crossing  $x=L$ and reaching $x=-L$ for the first time.
The mean first-passage time (MFPT) $\tf$ is subsequently calculated by averaging over all 
individual first-passage events.

\begin{figure*}[ht]
	\includegraphics[width=\textwidth]{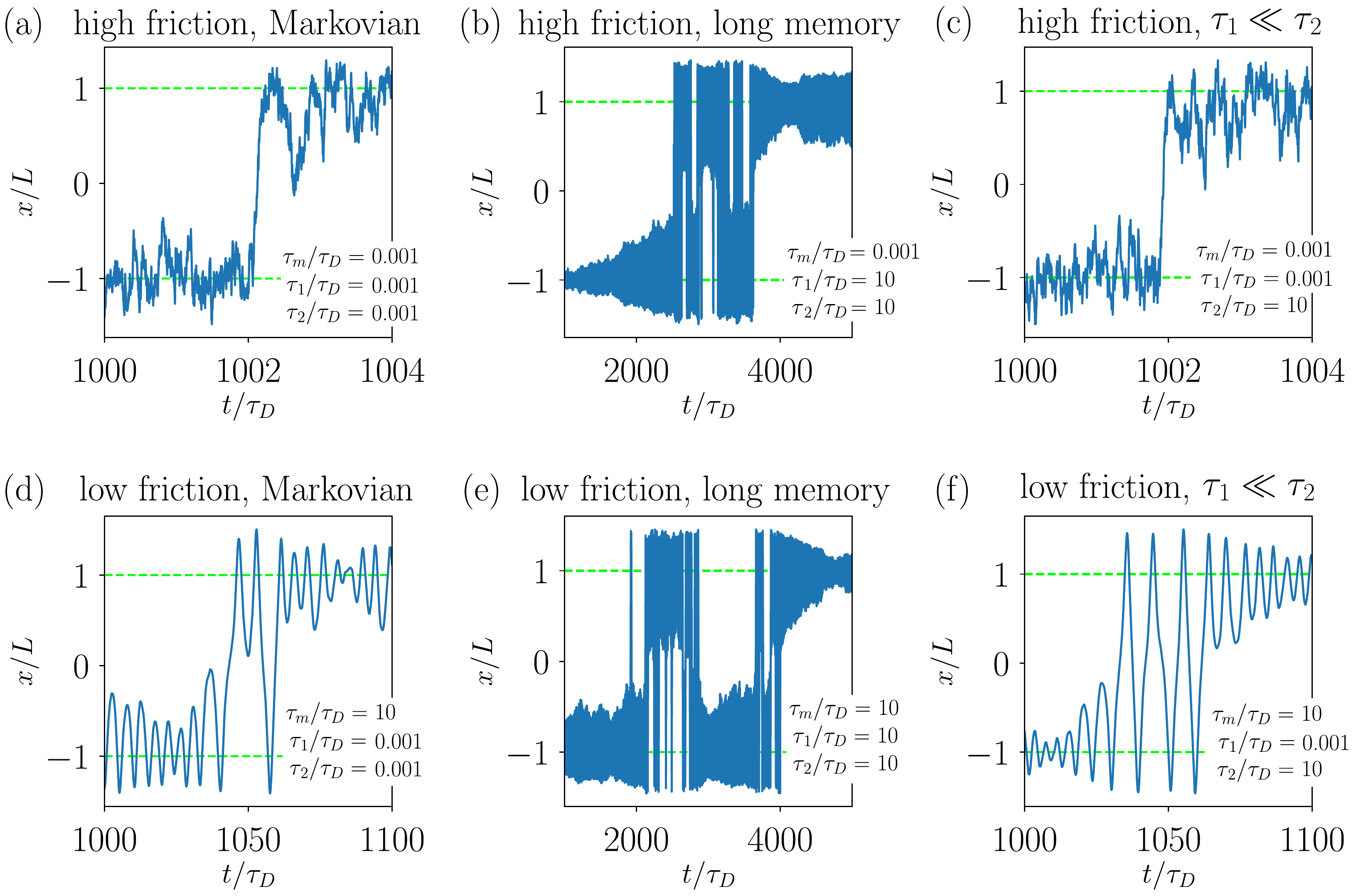}
\caption{
\label{fig:DoubleExp_Trajectories}
{Typical  simulation  trajectories  that display barrier-crossing events for fixed barrier height $\beta U_0 = 3$.}
Simulation parameters used are given in the legends.
The horizontal  dashed lines indicate the minima
 of the quartic potential Eq.~\eqref{eq:DoubleExp_QuarticPotential}, 
which is shown in Fig.~\ref{fig:DoubleExp_Potential} (a).
}
\end{figure*}

\begin{figure*}%[ht]
	\includegraphics[width=\textwidth]{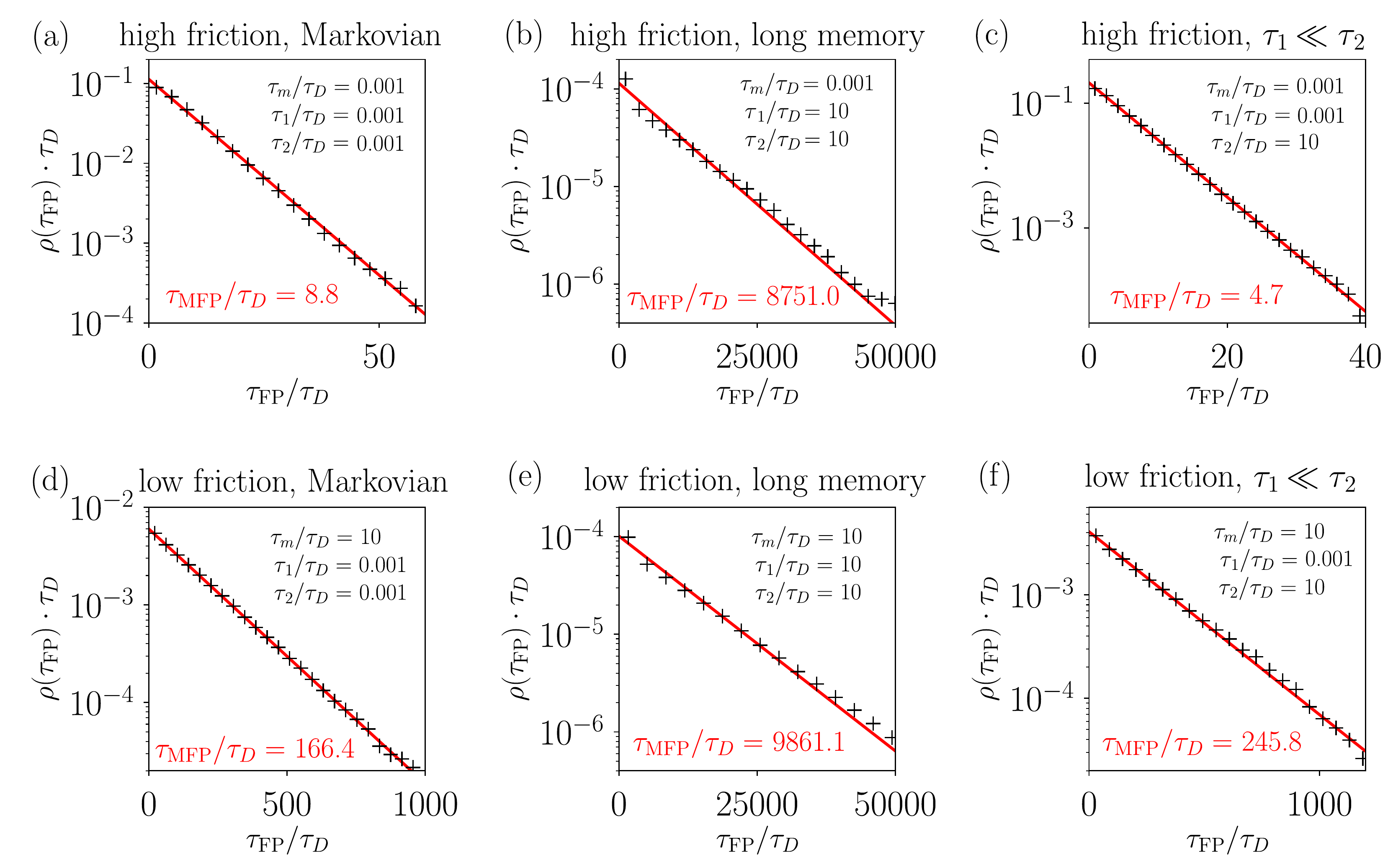}
\caption{
\label{fig:DoubleExp_FPT_dists}
{First-passage time (FPT) distributions $\rho(\tfp)$ for barrier crossing from one potential well  to the other.}
Simulation parameters are given in the legends.
For each subplot, first-passage events  are obtained from the 
numerical simulations as illustrated in Fig.~\ref{fig:DoubleExp_Potential} (b).
The resulting normalized probability distribution $\rho(\tau_{\mathrm{FP}})$ is shown as black crosses.
The MFPT value $\tf$ given in the plots is obtained by averaging over all first-passage events,
and used to plot an exponential distribution as defined in Eq.~\eqref{eq:DoubleExp_FPT_Exp},
shown as a solid line.
}
\end{figure*}

\section{Trajectories and first-passage distributions}

In Fig.~\ref{fig:DoubleExp_Trajectories} we show typical simulation trajectories that
illustrate how  the character of barrier-crossing events changes with varying mass, friction,
 and memory time parameters.
In Fig.~\ref{fig:DoubleExp_Trajectories} (a), (b), (d), (e) the two memory times are equal, $\tg \equiv \tgOne = \tgTwo$, 
so that the memory function  is in fact  single-exponential.
For  high friction and  short memory in Fig.~\ref{fig:DoubleExp_Trajectories} (a),
 i.e.~for $\tm/\td \ll 1$ and  $\tg/\td \ll 1$,
 the barrier crossing is diffusive, meaning that the particle fluctuates in a potential well for a long time 
 until a single barrier-crossing event occurs with a transition path time that is much shorter than 
 the mean first-passage time
 \cite{eaton2013,orland2017}.
In contrast, if the friction is low or if the memory time is long, i.e.~for
 $\tm/\td \gg 1$ or $\tg/\td \gg 1$,
the trajectories in  Fig.~\ref{fig:DoubleExp_Trajectories} (b), (d), (e)
 are characterized by  bursts of multiple barrier recrossings. In these cases the 
 barrier-crossing dynamics is dominated by the diffusive exchange of energy 
 between the particle and the thermal bath;
once the particle has acquired enough  energy to cross the barrier, 
 it oscillates back and forth between the two wells until  its energy
falls below the barrier energy again  \cite{melnikov_theory_1986,kappler_memory-induced_2018}.
Figure \ref{fig:DoubleExp_Trajectories} (c) shows a trajectory for high friction $\tm/\td = 0.001  \ll 1$
and different memory times  $\tgOne/\td = 0.001 \ll \tgTwo/\td = 10$.
The trajectory looks similar  to the  high-friction short-memory trajectory 
for   $\tm/\td = 0.001  $ and $\tgOne/\td = \tgTwo/\td = 0.001$  shown in Fig.~\ref{fig:DoubleExp_Trajectories} (a),
and is markedly different from the  trajectory 
for   $\tm/\td = 0.001  $ and $\tgOne/\td = \tgTwo/\td = 10$  displayed in Fig.~\ref{fig:DoubleExp_Trajectories} (b),
which is characterized by extended periods of multiple recrossings.
We conclude  that  a high-friction  trajectory that is governed by the sum of a slowly decaying 
 memory contribution with   $\tgTwo/\td = 10$ and a quickly decaying
memory contribution with   $\tgOne/\td = 0.001$ looks like a trajectory with only short-time memory,
meaning that the long-time memory contribution appears to be negligible compared to
 the short-time memory
contribution. 
Similarly, the trajectory for low friction $\tm/\td = 10 \gg 1$ and  different memory times 
$\tgOne/\td = 0.001 \ll \tgTwo/\td = 10$ shown 
in Fig.~\ref{fig:DoubleExp_Trajectories} (f) resembles more the low-friction Markovian trajectory 
 for  $\tm/\td = 10 $ and   $\tgOne/\td =  \tgTwo/\td = 0.001$ in  Fig.~\ref{fig:DoubleExp_Trajectories} (d)
than the low-friction long-memory trajectory 
 for  $\tm/\td = 10 $ and   $\tgOne/\td =  \tgTwo/\td = 10$
in Fig.~\ref{fig:DoubleExp_Trajectories} (e).
Again, we observe that
the slowly decaying memory contribution has a negligible effect on a low-friction  trajectory in the presence of a 
second memory contribution with a much shorter memory time.

The analysis of  first-passage-time (FPT)  distributions allows a more quantitative 
comparison of barrier-crossing statistics.
Fig.~\ref{fig:DoubleExp_FPT_dists} presents numerically obtained FPT distributions in a semi-logarithmic representation, 
calculated for the same parameters as for  the trajectories depicted in  Fig.~\ref{fig:DoubleExp_Trajectories}.
For all parameter combinations considered, 
 the  distributions  are well described  by a single-exponential distribution
\begin{equation}
	\label{eq:DoubleExp_FPT_Exp}
	\rho(\tfp) = \frac{1}{\tf}\exp({-\tfp/\tf}),
\end{equation}
shown as solid lines in Fig.~\ref{fig:DoubleExp_FPT_dists}.
Only for the long-memory cases shown in Fig.~\ref{fig:DoubleExp_FPT_dists} (b), (e) we see slight deviations
from an  exponential distribution for small $\tfp/\td$, which we attribute to the presence of
multiple recrossing events
seen in the trajectories shown in Fig.~\ref{fig:DoubleExp_Trajectories} (b), (e), but  which do not affect the MFPT significantly, 
as was previously demonstrated  in Ref.~\cite{kappler_memory-induced_2018}.
Both for high friction (upper row) and low friction (lower row), 
we see that if the two memory times are very different, as in Fig.~\ref{fig:DoubleExp_FPT_dists} (c), (f),  the resulting
FPT distribution is still single-exponential. Moreover, the 
 MFPTs in Fig.~\ref{fig:DoubleExp_FPT_dists} (c), (f)
   are  close to the corresponding MFPTs for single-exponential memory with the shorter memory time, 
 shown in Fig.~\ref{fig:DoubleExp_FPT_dists} (a), (d),
and  differ by  orders of magnitude from the single-exponential  MFPTs for  the longer memory time, 
 shown in Fig.~\ref{fig:DoubleExp_FPT_dists} (b), (e).
We conclude  that the barrier-crossing statistics 
in the presence of  bi-exponential memory is characterized by single-exponential FPT distributions 
which are dominated by the shorter memory time. 
This result is corroborated by a pole analysis of the positional autocorrelation function 
obtained  form the linearized generalized Langevin equation, 
see Appendix \ref{sec:DoubleExponential_PropagatorAnalysis}, where we show that indeed 
 for $\td \ll \tgOne \ll \tgTwo$, the particle motion  is dominated by $\tgOne$.

\begin{figure*}%[ht]
\includegraphics[width=0.95\textwidth]{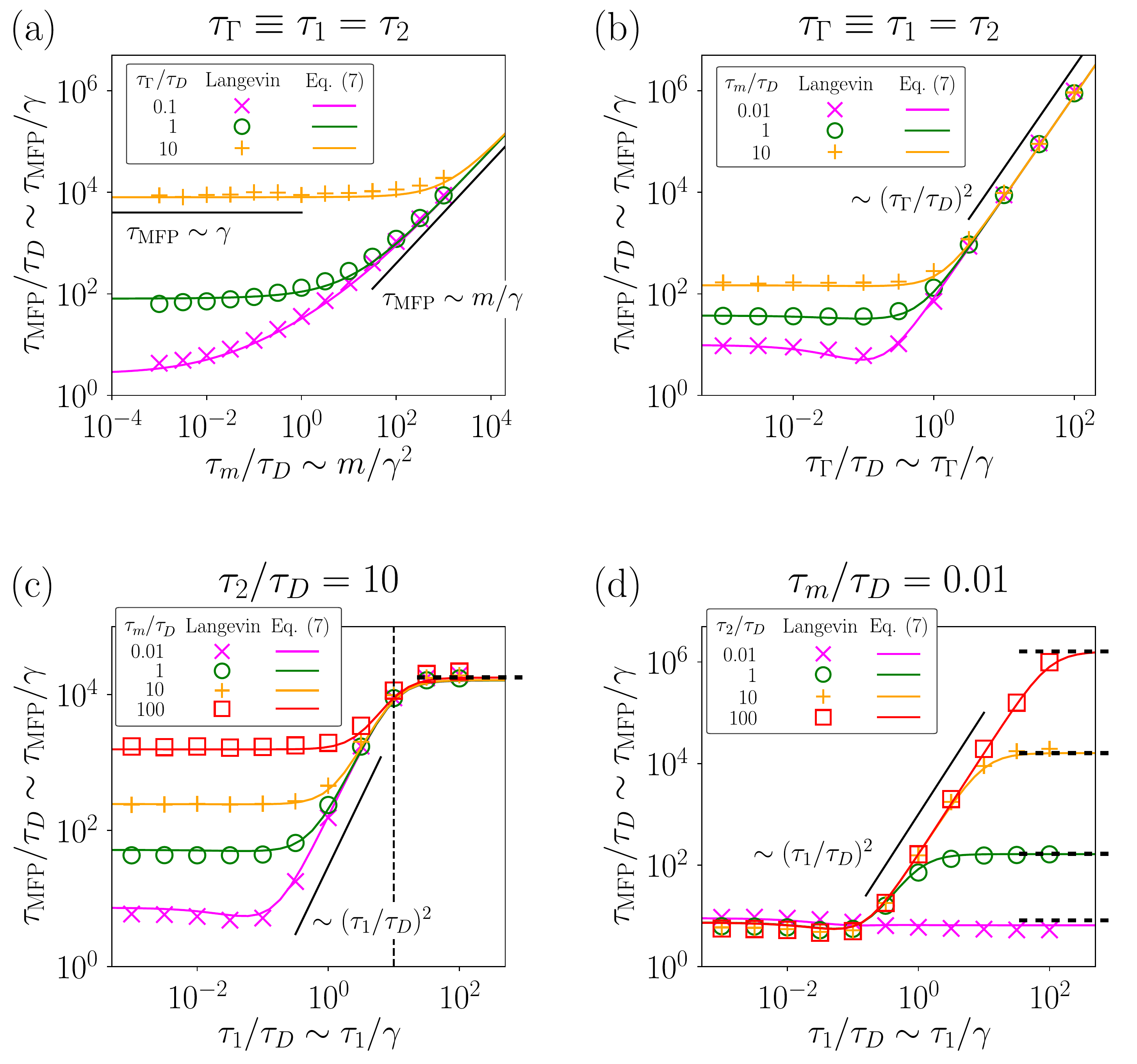}
\caption{
Throughout  the figure, colored symbols denote MFPT simulation results and  
colored lines represent the crossover formula Eq.~\eqref{eq:tf_multi},
the barrier height is fixed at  $\beta U_0 = 3$.
(a), (b) {MFPT results for single-exponential memory,} i.e., for $\tg \equiv \tgOne = \tgTwo$.
(a) The rescaled MFPT $\tf/\td$ is shown as a function of the rescaled inertial time
$\tm/\td$ for several values of the rescaled memory time $\tg/\td$.
The asymptotic power law scalings $\tf/\td \sim \tm/\td$ (Markovian low-friction regime) and
$\tf/\td \equiv$ const. (Markovian high-friction regime) are shown as black bars.
(b) The rescaled MFPT $\tf/\td$ is shown as a function of $\tg/\td$ for several values of $\tm/\td$.
The asymptotic power law scaling $\tf/\td \sim (\tg/\td)^2$  \cite{straub_non-markovian_1986,kappler_memory-induced_2018}
 is shown as a black bar.
(c) MFPT results for bi-exponential memory at fixed $\tgTwo/\td = 10$
 as function of $\tgOne/\td$ for several fixed values of $\tm/\td$.
The value at which $\tgOne = \tgTwo$ is indicated by a vertical dashed line.
The black bar indicates the intermediate scaling $\tf/\td \sim (\tgOne/\td)^2$, 
the  horizontal dashed black line to the right denotes
the prediction  from Eq.~\eqref{eq:tf_multi} in  the limit  $\gamma_1=0$.
(d)  MFPT results for bi-exponential memory at fixed $\tm/\td = 0.01$
 as function of $\tgOne/\td$ for several  values of $\tgTwo/\td$.
The black bar indicates the intermediate scaling $\tf/\td \sim (\tgOne/\td)^2$,
the  horizontal dashed black lines  to the right denote
the predictions  from Eq.~\eqref{eq:tf_multi} in  the limit  $\gamma_1=0$.
}
\label{fig:MFPTs}
\end{figure*}

\section{MFPT results for single-exponential memory and  general crossover formula  }

We first consider the symmetric case
$\tg \equiv \tgOne = \tgTwo$ where the memory kernel Eq.~\eqref{eq:DoubleExp_DoubleExpMemKern} 
reduces to a single exponential.
For fixed barrier height $\beta U_0 = 3 $ one only has two parameters, namely the 
inertial time  $\tm/\td$ and the memory time $\tg/\td$, both rescaled by the diffusion time.
\textcolor{black}{The dependence of the MFPT on $\beta U_0 $ has
 previously been considered \cite{kappler_memory-induced_2018},
and for the present case is discussed in Appendix \ref{app:BarrierHeight}.}
%.
%
In Fig.~\ref{fig:MFPTs} (a) we show  MFPTs  from simulations for various fixed values of $\tg/\td$ as a function
of $\tm/\td$. 
As can be most clearly seen for the $\tg/\td =$ 1 data, for $\tm/\td \ll 1$ the MFPT  $\tf/\td$
 becomes independent of $\tm/\td$; this is the classical Kramers high-friction regime where
  $\tf \sim \gamma$ \cite{kramers_brownian_1940,melnikov_theory_1986}.
In the opposite limit $\tm/\td \gg 1$, the MFPT scales as $\tf \sim m/\gamma$, indicating the 
Kramers low-friction (energy diffusion) limit \cite{kramers_brownian_1940,melnikov_theory_1986}.
In Fig.~\ref{fig:MFPTs} (b) we show simulated  MFPTs for various fixed values of $\tm/\td$ as a function
of $\tg/\td$.
While for long memory, $(\tg/\td)^2 \gg \max\{1, \tm/\td\}$, the MFPT scales as $\tf \sim \tg^2$ \cite{straub_non-markovian_1986,kappler_memory-induced_2018}, in the Markovian regime 
 $\tg/\td \ll 1$ the MFPT is independent of $\tg/\td$.
For high friction  $\tm/\td \ll 1$, an intermediate regime where memory accelerates barrier crossing
compared to the Markovian limit $\tg/\td \ll 1$,
centered around  $\tg/\td \approx 0.1$, can be observed
 \cite{kappler_memory-induced_2018}.
We thus see that memory can in the high friction case either accelerate or slow down the barrier-crossing dynamics,
depending on the memory time. 
For long memory the barrier-crossing time grows  quadratically with the memory time both for  high friction 
and   low friction. This asymptotic  regime is intriguing, as it demonstrates  that memory modifies
the barrier-crossing dynamics, and in particular the mean first-passage time $\tf$, even when the memory time
$\tg$ is much shorter than $\tf$. 

We now present a crossover  formula 
for the MFPT for  a general multi-exponential memory kernel,
\begin{align}
%	\label{eq:DoubleExp_DoubleExpMemKern}
	\Gamma(t) &= \sum_{i=1}^N \frac{\gamma_i}{\tgi} \exp\left(-{|t|}/{\tgi}\right) %\\
\end{align}
parametrized by friction coefficients $\gamma_i$ and memory times $\tgi$,
and define the friction coefficient as $ \gamma = \sum_{i=1}^N  \gamma_i$.
Inspired by  the linear dependence of  the reaction time on friction  in the 
overdamped limit \cite{kramers_brownian_1940}
and the linear dependence of  the reaction rate on the memory kernel   in the 
energy-diffusion limit \cite{talkner_transition_1988}, we construct  a heuristic
crossover formula for the MFPT  as
\begin{equation}
	\label{eq:tf_multi}
	\tf = \sum_{i=1}^n \tfovi + \left( \sum_{i=1}^n  1/\tfedi \right)^{-1},
\end{equation}
where the overdamped MFPT contribution  $\tfovi$ and the energy-diffusion MFPT contribution  $\tfedi$ are given by 
\begin{align}
	\label{eq:tfov_single}
	\frac{\tfovi}{\td} &= \frac{\gamma_i}{\gamma} \frac{e^{\beta U_0}}{\beta U_0} \left[ \frac{\pi}{2 \sqrt{2}} \frac{1}{1+ 10 \beta U_0 \tgi/\td} +\sqrt{\beta U_0 \frac{\tm}{\td}} ~\right],\\
	\frac{\tfedi}{\td} &= \frac{\gamma}{\gamma_i}\frac{e^{\beta U_0}}{\beta U_0} \left[ \frac{\tm}{\td} + 4 \beta U_0 \left( \frac{\tgi}{\td}\right)^2  + \sqrt{\beta U_0 \frac{\tm}{\td}} ~\right].
		\label{eq:tfed_single}
\end{align}

The first sum in Eq.~\eqref{eq:tf_multi} reflects  that in the  memoryless high-friction scenario 
 the MFPT scales as  $\tf \sim \gamma = \sum_{i} \gamma_i$, as follows  from 
the   Kramers theory  in the high-friction  limit \cite{kramers_brownian_1940}.
The inverse additivity of the individual  MFPT contributions in the energy-diffusion regime,
the second sum in Eq.~\eqref{eq:tf_multi},
is derived in Appendix \ref{app:inverse_additivity}.
{The additivity of the overdamped- and energy-diffusion MFPT contributions
in Eq.~\eqref{eq:tf_multi} was previously used to construct 
similar crossover formulas \cite{weiss_1984,nitzan_1984,berne_1985}.}
Our crossover formula for the MFPT  obeys an important symmetry: 
In the case of single-exponential memory, i.e.~if all $\tgi$ are equal, Eq.~\eqref{eq:tf_multi}  
only depends on 
the $\gamma_i$ only via their sum $\gamma$, as it should. In this limit our expression slightly deviates from 
our previous  single-exponential crossover  formula \cite{kappler_memory-induced_2018}:
First, the prefactor $\pi/(2\sqrt{2}) \approx 1.11$ in Eq.~\eqref{eq:tfov_single} replaces a factor $1$ in Ref.~\cite{kappler_memory-induced_2018},
so that Eq.~\eqref{eq:tfov_single}  reproduces the overdamped Kramers limit $\tm/\td \ll 1$, $\tg/\td \ll 1$ \cite{kramers_brownian_1940} exactly.
Second, the prefactor $4$ in the second term in Eq.~\eqref{eq:tfed_single} replaces a factor $e \approx 2.72$ in Ref.~\cite{kappler_memory-induced_2018}, which leads to improved matching with simulated MFPTs in 
the long-memory regime $(\tg/\td)^2 \gg \max\{1, \tm/\td\}$.

Equation \eqref{eq:tf_multi} is  in Fig.~\ref{fig:MFPTs} (a), (b) included as colored solid lines
 and agrees with the numerical data very well. In particular, it reproduces 
all  asymptotic  scaling regimes of the single-exponential scenario, namely:
i) the memoryless high-friction Kramers regime $\tg/\td \ll 1$ and $\tm/\td \ll 1$, 
given by the first term in Eq.~\eqref{eq:tfov_single} for
$\tg/\td = 0$ and observed for $\tm/\td \ll 1$ in Fig.~\ref{fig:MFPTs} (a);
ii) the memoryless low-friction regime $\tg/\td \ll 1$ and $\tm/\td \gg 1$, given by the first term in Eq.~\eqref{eq:tfed_single}
and observed for $\tm/\td \gg 1$ in Fig.~\ref{fig:MFPTs} (a);
 iii) the long-memory regime $(\tg/\td)^2 \gg \max\{1, \tm/\td\}$, given by the
 second term in Eq.~\eqref{eq:tfed_single}
 and observed for $\tg/\td \gg 1$ in Fig.~\ref{fig:MFPTs} (b).
The  last terms in Eqs.~\eqref{eq:tfov_single} and \eqref{eq:tfed_single} are 
included to improve the crossover between the overdamped and energy-diffusion regimes
and are fitted to the simulation MFPT data.

\textcolor{black}{
In Appendix \ref{app:BarrierHeight}, Fig.~\ref{fig:ComparisonU} (a), 
we compare the crossover formula Eq.~\eqref{eq:tf_multi} with
simulation results for varying $\beta U_0$ in the symmetric case $\tgOne = \tgTwo$;
there, we find good agreement between the crossover formula and numerical results
for $\beta U_0 \gtrsim 2$.}

\section{MFPT results  for bi-exponential memory}

In Fig.~\ref{fig:MFPTs} (c), (d) we compare simulation results  for $\tf$ with the crossover formula
 Eq.~\eqref{eq:tf_multi} for the asymmetric scenario where the  bi-exponential  memory times 
 $\tgOne$  and  $\tgTwo$
 are unequal.
Figure \ref{fig:MFPTs} (c) shows $\tf/\td$ as function of $\tgOne/\td$ 
for fixed $\tgTwo/\td = 10$ and several values of $\tm/\td$.
Throughout Fig.~\ref{fig:MFPTs} (c) the crossover  formula 
Eq.~\eqref{eq:tf_multi}, denoted by colored solid lines, describes the simulation  results
very well.
We see that for $\tgOne \ll \tgTwo$, 
i.e.~to the left of the vertical dashed line that denotes $\tgOne = \tgTwo$, 
the MFPT  behaves very  similarly  to the single-exponential MFPT results shown in  Fig.~\ref{fig:MFPTs} (b).
As $\tgOne \gtrsim \tgTwo$, i.e.~to the right of the  vertical dashed line, 
the MFPT becomes independent of $\tgOne$
and takes on the value which is obtained in the absence of the more slowly  decaying exponential memory component,
as indicated by the horizontal dashed black  line which follows from Eq.~\eqref{eq:tf_multi} by taking the limit  $\gamma_1=0$.
\textcolor{black}{In agreement with the behavior of the particle  trajectories and the FPT distributions discussed before,
the MFPT only depends on the shorter of the two memory times 
if at least one of the two memory times $\tgOne$, $\tgTwo$ is larger than $\td$.}

Figure \ref{fig:MFPTs} (d) shows $\tf/\td$ as function of $\tgOne/\td$
for fixed $\tm/\td = 10$ and several  values of $\tgTwo/\td$ as indicated in the legend.
Also in this plot, the crossover  formula Eq.~\eqref{eq:tf_multi} describes the simulation data very 
accurately.
For $\tgOne \ll \tgTwo$ we again obtain   behavior
reminiscent of the single-exponential results shown in Fig.~\ref{fig:MFPTs} (b).
As $\tgOne \gtrsim \tgTwo$ and 
$ \tgOne > \td$, $\tf/\td$ saturates at a value that only depends on
$\tm/\td$ and $\tgTwo/\td$, while the value of $\tgOne$ becomes irrelevant.
The horizontal dashed black lines to the right of the figure
denote the predictions  from Eq.~\eqref{eq:tf_multi} in the limit  $\gamma_1=0$.
For $\tgTwo/\td = 0.01$ (magenta crosses), Fig.~\ref{fig:MFPTs} (d) contains
no regime where $\td < \tgOne \ll \tgTwo$,
so that $\tf$ is almost independent of $\tgOne/\td$ throughout.

In Appendix \ref{app:GlobalComparison} we compare our  crossover formula to simulation 
data for a wider range of the parameters \textcolor{black}{$\tm/\td$, $\tgOne/\td$, $\tgTwo/\td$ at
 barrier height $\beta U_0 = 3$},
  confirming that 
Eq.~\eqref{eq:tf_multi} globally describes bi-exponential barrier crossing very well and that $\tf$ 
is dominated by the shorter memory time for all parameter values.
\textcolor{black}{In Appendix \ref{app:BarrierHeight} we show that our crossover formula Eq.~\eqref{eq:tf_multi}
 agrees with simulations also for larger barrier heights $\beta U_0$.}
As demonstrated in  Appendix~\ref{app:GH},
 Grote-Hynes (GH) theory \cite{grote_stable_1980} only describes the simulation  data in the triple-limit of
 high-friction $\tm/\td \ll 1$ and short memory $\tgOne/\td \ll 1$, $\tgTwo/\td \ll 1$,
\textcolor{black}{which is why we do not compare our results with predictions of GH theory in the main text.}

\begin{figure*}[ht]
\includegraphics[width=0.9\textwidth]{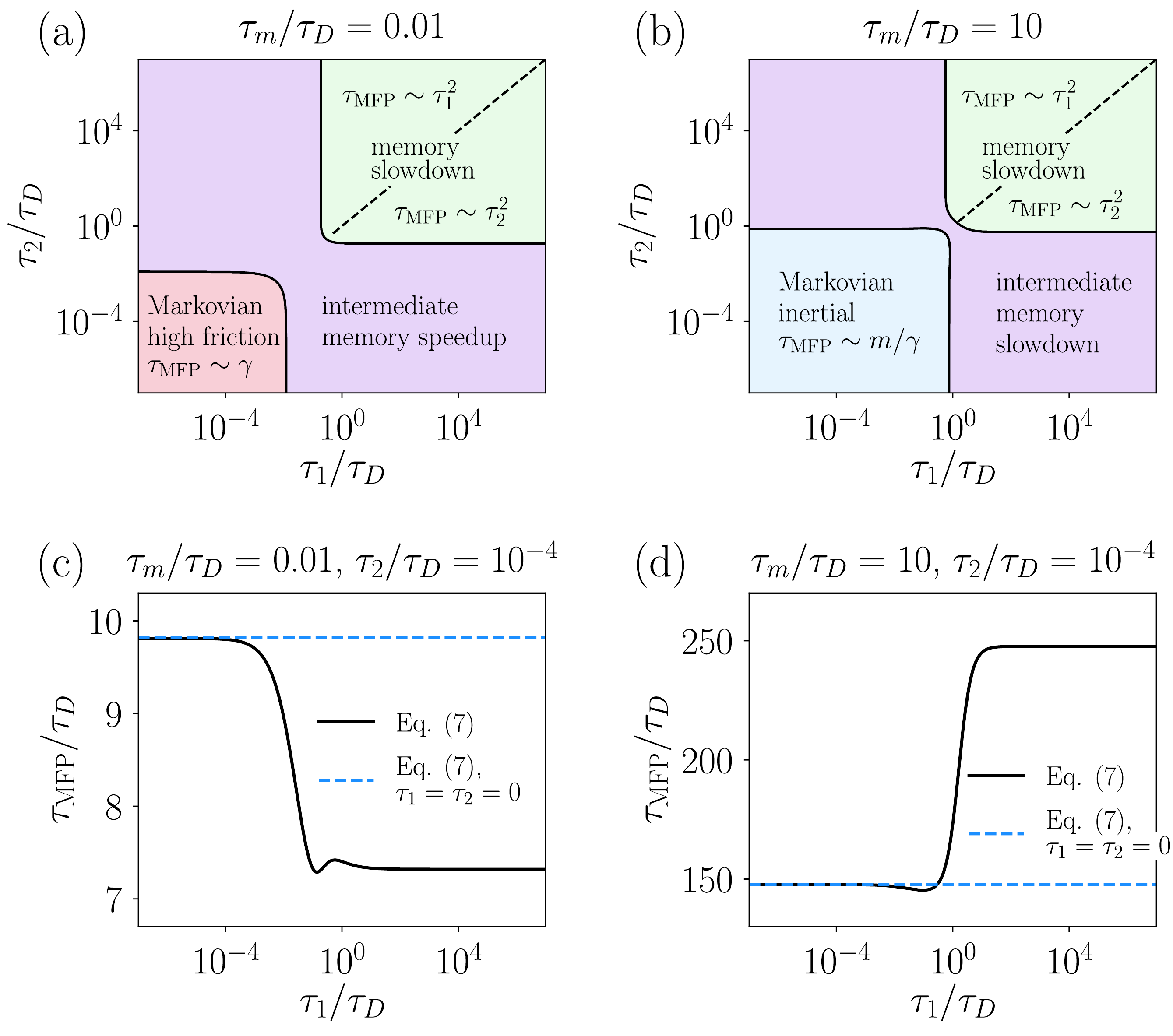}
\caption{
(a), (b) Scaling diagrams for the dependence of the  MFPT $\tf$ on 
the rescaled memory times $\tgOne/\td$ and $\tgTwo/\td$ for
(a) high friction $\tm/\td = 0.01$ and (b) low friction $\tm/\td = 10$.
In  (a), the transition from the Markovian high-friction regime to the intermediate memory speedup regime is defined 
by the location where $\tf$ is smaller by 10\% than the Markovian limit  $\tgOne = \tgTwo = 0$
(based on  the crossover  formula Eq.~\eqref{eq:tf_multi}).
The transition from the intermediate memory speedup regime  to the asymptotic 
memory slowdown regime is defined by the location where  $\tf$ equals  the Markovian limit.
In  (b), the transition from the Markovian low-friction regime to the intermediate memory slowdown regime is defined 
by the location where $\tf$ is larger  by 10\% than the Markovian limit  $\tgOne = \tgTwo = 0$
(based on  the crossover  formula Eq.~\eqref{eq:tf_multi}).
The transition from the intermediate memory slowdown  regime  to the asymptotic 
memory slowdown regime is defined by the location where  $\tf$ is twice as large as in  the Markovian limit.
The dashed diagonal  lines in (a) and (b)  in the asymptotic memory slowdown regime indicate
 the crossover from $\tgOne$-dominated
barrier crossing for $\tgOne \ll \tgTwo$, to $\tgTwo$-dominated barrier crossing for $\tgTwo \ll \tgOne$.
(c), (d) 
The  rescaled MFPT  $\tf/\td$  according to  Eq.~\eqref{eq:tf_multi} is shown as a function of $\tgOne/\td$ for fixed
$\tgTwo/\td = 10^{-4}$ and (a) $\tm/\td = 0.01$, and (b) $\tm/\td = 10$.
The dashed line denotes the Markovian limit, obtained from Eq.~\eqref{eq:tf_multi} by setting $\tgOne = \tgTwo = 0$.
All data is obtained for a fixed barrier height of  $\beta U_0 = 3$.
\label{fig:ScalingDiagrams}
}
\end{figure*}

The  dependence of the MFPT on  the memory times $\tgOne$ and $\tgTwo$ is 
summarized in  Fig.~\ref{fig:ScalingDiagrams} in terms of  scaling diagrams for both high and low friction. 
For high friction, $\tm/\td = 0.01$, we see in Fig.~\ref{fig:ScalingDiagrams} (a) that 
 the Markovian high-friction regime, which corresponds to  the Kramers high-friction regime where
 $\tf \sim \gamma$ \cite{kramers_brownian_1940},
 is obtained  when both  $\tgOne/\td$ and  $\tgTwo/\td $ become small.
For $\tgOne/\td>1$ and  $\tgTwo/\td >1$
the  memory slowdown regime is reached 
where the MFPT increases  quadratically with the memory time. 
This regime is divided along the diagonal, 
since  the shorter memory time dominates the   barrier-crossing kinetics, such that for
 $\tgOne >  \tgTwo $ one finds  $\tf \sim \tgTwo^2$ while for 
 $\tgOne <  \tgTwo $ one finds  $\tf \sim \tgOne^2$.
The Markovian and memory slowdown asymptotic regimes  are separated by an intermediate 
memory speedup regime, where barrier
crossing is slightly accelerated by the presence of memory as compared to the Markovian limit.
Along the diagonal $\tgOne = \tgTwo$, the  intermediate memory speedup and the asymptotic memory slowdown
regimes are illustrated in Fig.~\ref{fig:MFPTs} (b);
parallel to the $\tgOne$-axis, the memory speedup is illustrated in Fig.~\ref{fig:ScalingDiagrams} (c)
by a plot of   $\tf$  for constant $\tgTwo/\td =10^{-4}$.

For low friction, $\tm/\td = 10$, shown in Fig.~\ref{fig:ScalingDiagrams} (b), the Markovian
regime is obtained for $\tgOne/\td < 1$ and  $\tgTwo/\td < 1$.
For $\tgOne/\td   \gg \max\{1,\sqrt{\tm/\td}\} = \sqrt{10}$  and  $\tgTwo/\td  \gg \max\{1,\sqrt{\tm/\td}\} = \sqrt{10}$
the  memory slowdown regime is reached  where $\tf \sim \tgTwo^2$ or  $\tf \sim \tgOne^2$,
depending on which memory time is smaller.
In between  these two asymptotic regimes, we find an intermediate regime
where the barrier crossing is slowed down compared to the Markovian limit 
but the MFPT does not display the quadratic memory-time dependence, which 
is illustrated  in Fig.~\ref{fig:ScalingDiagrams} (d)
by a plot of   $\tf$  for constant $\tgTwo/\td =10^{-4}$.
A  plot of $\tf$ along the diagonal for $\tgOne = \tgTwo$ is shown in 
 Fig.~\ref{fig:MFPTs} (b).

There are different levels on which one can  rationalize and  intuitively understand
the main result obtained here, namely that it is the memory contribution with the
shorter decay time that dominates the barrier-crossing kinetics in the non-Markovian limit. 
On a formal level, the linear  pole analysis of the positional autocorrelation function 
in Appendix \ref{sec:DoubleExponential_PropagatorAnalysis} 
 shows that    for $\td < \tgOne \ll \tgTwo$,
 the memory contribution that depends on $\tgTwo$ can be neglected, 
  so that   the particle motion  only depends on $\tgOne$.

An alternative viewpoint is obtained by  mapping of the non-Markovian 
one-dimensional Langevin equation onto a  system of $N$ coupled Markovian 
degrees of freedom, see Appendix \ref{app:TransformingToMarkovian}.
In  this picture, 
each exponential contribution  to the memory function $\Gamma(t) $ can be viewed
 as an independent heat bath with  relaxation time $\tgi$, with which the particle
 exchanges energy, see Appendix \ref{app:TransformingToMarkovian}.
Accordingly, the inverse additivity of the MFPTs of each exponential memory contribution
 in the energy-diffusion regime, 
represented by the second term in Eq.~\eqref{eq:tf_multi},
means that $\tf$ is  dominated  by the heat bath with the shortest relaxation time, i.e.~by the heat bath
which is the fastest in providing the particle with the energy needed for barrier crossing,
which (for identical $\gamma_i$) is the heat bath with the smallest $\tgi$.
In conclusion, for bi-exponential memory, the shorter of the two
 memory times dominates the barrier-crossing time simply because it
provides the barrier-crossing energy faster.

Yet another way of understanding the dominance of the memory contribution with the shorter decay time
is obtained by   considering the memory
integral in Eq.~\eqref{eq:DoubleExp_LangevinDoubleExp}
for a bi-exponential  kernel  given by Eq.~\eqref{eq:DoubleExp_DoubleExpMemKern}.
The particle will undergo many oscillations (for low friction) or random fluctuations (for high friction) 
within a well before attempting to cross the barrier. 
If the memory times are longer than the intrinsic   relaxation  time of the particle
(which can be the intrinsic diffusion time $\td$ or the oscillation period depending on
whether one is in the high-friction or the low-friction regime),
the convolution of the particle velocity history  
with the exponential memory contribution with the longer memory time will give a smaller friction contribution 
than the convolution with the exponential memory contribution with the shorter memory time (again assuming
 equal memory amplitudes $\gamma_1$ and $\gamma_2$), simply because the mean of the particle velocity  is zero. 
This shows  directly that  memory effects in confinement tend to be dominated by the shorter memory contributions.

\begin{figure*}[ht]
	\centering
	\includegraphics[width=0.9\textwidth]{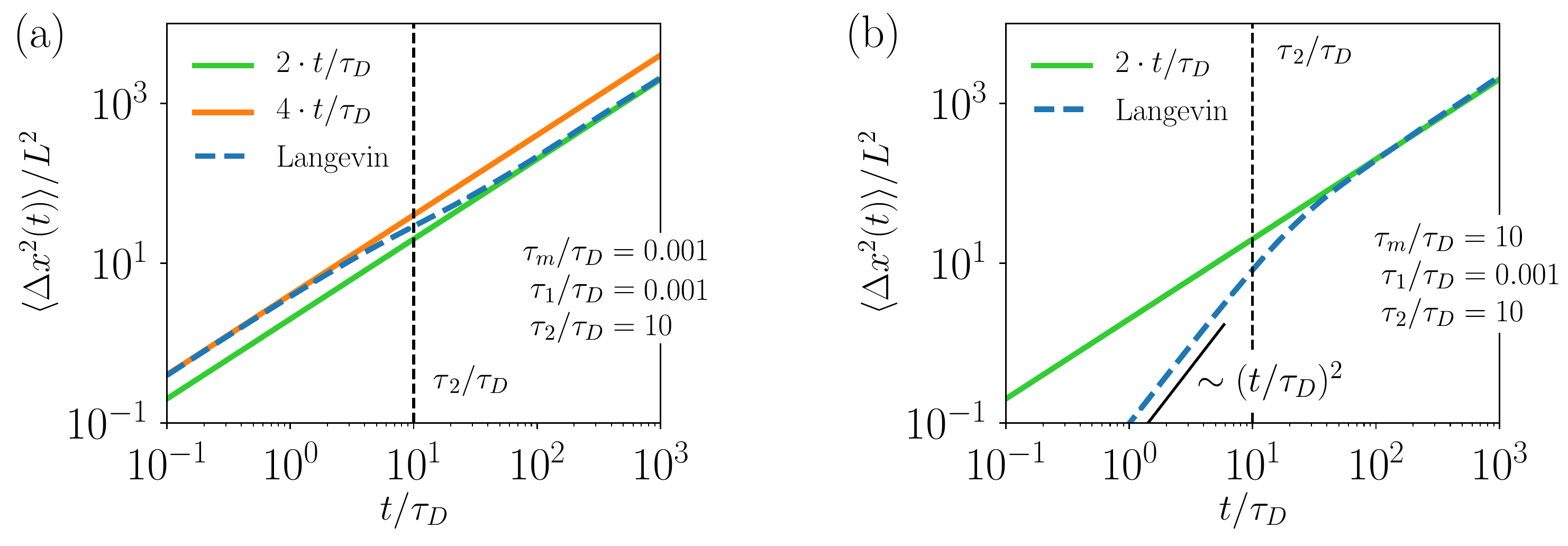}
\caption{
Comparison of the simulated mean-squared displacement (MSD), shown as dashed blue line (consisting of 8 data points per decade),  with asymptotic predictions 
in the absence of an  external potential for $U(x)=0$.
(a) Results for high friction 
$\tm/\td = 0.001$ and memory times $\tgOne/\td = 0.001$ and $\tgTwo/\td = 10$, the same parameters as used in
Fig.~\ref{fig:DoubleExp_Trajectories} (c).
The solid lines denote the expected MSD for diffusion with friction coefficients $\gamma/2$ (orange line) and
$\gamma$ (green line).
The  vertical dashed line denotes $\tgTwo/\td$.
{(b)} Results for low friction $\tm/\td = 10$ and memory times  $\tgOne/\td = 0.001$ and $\tgTwo/\td = 10$,
the same parameters as used in Fig.~\ref{fig:DoubleExp_Trajectories} (f).
The solid green  line denotes the expected MSD for diffusion with friction coefficient $\gamma$.
The black bar  indicates the power law $\langle \Delta x^2(t) \rangle\sim t^2$ expected for ballistic motion, 
the  vertical dashed line denotes $\tgTwo/\td$. 
\textcolor{black}{Note that for free diffusion there exist analytical expressions for both the
 Laplace- and Fourier-transform of the MSD \cite{grabert_1988,netz_2019}, 
 so that Fig.~\ref{fig:DoubleExp_MSDs} could have been
 generated without invoking numerical simulations.}
}
\label{fig:DoubleExp_MSDs}
\end{figure*}

\section{Free diffusion case }

While barrier-crossing dynamics is dominated by the memory contribution with the shorter memory time
in the non-Markovian limit,
in the absence of an external potential, i.e.~for $U(x)=0$, the long-time diffusive behavior is 
always expected to be
determined by  the total friction $\gamma = \gamma_1 + \gamma_2$.
In Fig.~\ref{fig:DoubleExp_MSDs} we show mean-squared displacements (MSDs)
\begin{equation}
	\langle \Delta x^2(t) \rangle = \left\langle \left[x(t)-x(0)\right]^2 \right\rangle
\end{equation}
 obtained from simulations without an external potential, $U(x)=0$, and unequal memory times
 $\tgOne/\td = 0.001$, $\tgTwo/\td = 10$,
 as blue dashed lines.
These  parameters used are the same as in Fig.~\ref{fig:DoubleExp_Trajectories} (c), (f).
Figure \ref{fig:DoubleExp_MSDs} (a) shows the high-friction scenario $\tm/\td = 0.001$.
For times  $t \gg \tgTwo$ the MSD is well described  by 
\begin{equation}
	\label{eq:DoubleExp_MSD_FullDiff}
	\langle \Delta x^2(t) \rangle  = 2 \frac{1}{\beta \gamma} t= 2D t ,
\end{equation}
which in Fig. \ref{fig:DoubleExp_MSDs} (a) is denoted by a green solid line, and which is 
the expected result for the MSD of a Brownian particle with friction coefficient $\gamma$
and where we used the  Einstein relation to define the diffusion constant as  $D =1/(\beta \gamma)$.
For shorter  times $\tgOne < t < \tgTwo$, only the shorter memory contribution,
characterized by $\tgOne$, contributes to the friction  and the simulated  MSD is well 
described  by
\begin{equation}
%	\label{eq:DoubleExp_MSD_FullDiff}
	\langle \Delta x^2(t) \rangle  = 2 \frac{1}{\beta \gamma_1} t= 4D t ,
\end{equation}
which in Fig. \ref{fig:DoubleExp_MSDs} (a) is denoted by an orange solid line.

For the low-friction scenario $ \tm/\td =10$ depicted in Fig.~\ref{fig:DoubleExp_MSDs} (b),  the 
inertial regime  where $\langle \Delta x^2(t) \rangle   \sim t^2/(m \beta) $ appears 
 for $t \ll \tm$. For  $t \gtrsim \tm$ the MSD  exhibits diffusive behavior  determined by 
the full friction coefficient $\gamma$, i.e.  
$\langle \Delta x^2(t) \rangle  = 2t /  (\beta \gamma) = 2D t$,
since for $t \gg \tm$  we simultaneously fulfill  $t \gg \tgOne$ and $t\gg \tgTwo$.

We conclude that while barrier-crossing kinetics  is completely dominated
by the shorter memory time if the particle relaxation time is shorter than the longer memory time,
 the long-time  free diffusion behavior in the presence of
 bi-exponential memory  is  determined by  the sum of both 
 exponential memory contributions. 
This simply reflects that the particle relaxation time
 diverges in the free-diffusion scenario. 

\section{Conclusions}

\textcolor{black}{
 Our   Langevin simulations  in the presence of a bi-exponential memory function
 show  that if at least one of the two memory times $\tgOne$, $\tgTwo$ is larger than $\td$, 
then the barrier-crossing time $\tf$ is  dominated by the shorter 
of the two memory times.}
The simulation results are obtained  for the restricted case where the  exponential memory
contributions have the same integral weight. 
Based on asymptotic matching and  symmetry considerations, 
we construct a crossover formula for  $\tf$
which describes our simulation results for all parameters very well.
This formula applies to general 
 multi-exponential memory function  with arbitrary weights and decay times and 
corroborates that the  shortest memory time is expected to dominate $\tf$ in the non-Markovian limit.

 Our results have a number of important consequences:
Usually it is assumed that  the slow degrees of freedom of a system   dominate rare events
and in particular  the MFPT  $\tf$;  for the case of barrier crossing in the presence of 
memory functions with different time scales, our results show that  this 
time scale separation principle  is violated.
Our findings suggest that instead of  keeping only  the slowest  degrees of freedom,
which is conventionally done in coarse graining procedures,
one also needs to  keep the fastest orthogonal degrees of freedom if  non-Markovian effects are important. 
Of course, the choice of  the degrees of freedom that should be included  requires 
good understanding of all  relevant time scales. 

Non-Markovian effects have been demonstrated to be important for fast molecular transition such as 
the dihedral barrier dynamics of butane in   solvents \cite{de_sancho_molecular_2014,daldrop_butane_2018}. 
The effect of the different memory times of the multi-scale memory kernel 
that have been extracted from simulations will have to be examined in future work. Likewise,
protein dynamics has been suggested to be subject to memory effects in a number of works 
\cite{plotkin_non-markovian_1998,Makarov2018}. 
Here our  crossover  formula Eq.~\eqref{eq:tf_multi} will be useful,
because it  fills the gap between theoretical works on 
non-Markovian barrier crossing, which are typically confined to single-exponential memory \cite{straub_non-markovian_1986,pollak_theory_1989,ianconescu_study_2015},
and real physical systems, where orthogonal degrees of freedom 
typically feature  several  relaxation time scales \cite{rey_dynamical_1992,tolokh_prediction_2002,de_sancho_molecular_2014,lesnicki_molecular_2016,gottwald_parametrizing_2015,daldrop_external_2017,daldrop_butane_2018}.
Another field of application are chemical reactions in solvents, such as proton transfer reactions in water \cite{marx1999,marx2006}.
In such reactions, the friction is presumably rather small, such that the low-friction regime,
characterized by $ \tm/\td > 1$, is relevant. 
But also in the low-friction limit, our results show that for multi-scale memory   
the smaller memory time becomes 
dominant in the non-Markovian limit.

\begin{acknowledgments}
 Financial support from the
Deutsche Forschungsgemeinschaft (DFG)   via grant  SFB 1114 is
acknowledged.
\end{acknowledgments}

\appendix

\section{Dimensionless form of the GLE}
\label{app:DimensionlessGLE}

We consider the one-dimensional generalized Langevin equation
\begin{equation}
	\label{eq:DoubleExp_Dimensionless_Derivation_LangevinMultiExp}
	m \,\ddot{x}(t) = - \int_0^{t} \Gamma(t-t') \dot{x}(t')\,\mathrm{d}t' -  U'(x(t)) + \noise(t),
\end{equation}
where $m$ is the mass of the particle, $x$ its position, 
$\Gamma(t)$ the memory kernel and
 $U'(x)$ denotes the derivative of the external potential $U(x)$.
The random force $\noise(t)$ is Gaussian with zero mean, $\langle \noise(t) \rangle = 0$,
and obeys the generalized fluctuation-dissipation theorem (FDT)
\begin{align}
	\label{eq:DoubleExp_Dimensionless_Derivation_FDT}
	\beta \langle \noise(t)\noise(t') \rangle &=   \Gamma(|t-t'|).
\end{align}
We consider a multi-exponential memory kernel with friction coefficients $\gamma_i$ and memory times $\tgi$,
\begin{equation}
	\Gamma(t) = \sum_{i=1}^{N} \frac{\gamma_i}{\tgi}e^{-|t|/\tgi},
\end{equation}
and define the total friction 
%\begin{equation} 
%	\label{eq:DoubleExp_Dimensionless_Derivation_MemKern}
	$\gamma := \sum_{i=1}^N \gamma_i$.
%\end{equation}
Defining a dimensionless position $\xDL = x /L$ and a rescaled time
 $t = \tDL \tau_{D}$ and multiplying 
Eq.~\eqref{eq:DoubleExp_Dimensionless_Derivation_LangevinMultiExp}  by $\beta L$, we obtain
\begin{widetext}
\begin{equation}
	\label{eq:DoubleExp_Dimensionless_Langevin_LangevinMultiExpDL}
	\frac{\tm}{\td} \ddotxDL(\tDL) = -\sum_{i=1}^N \frac{\gamma_i}{\gamma} \frac{\td}{\tgi} \int_0^{\tDL}\mathrm{d}\tDL' \exp\left[-\frac{\td}{\tgi}\left( \tDL-\tDL'\right) \right] \dotxDL(\tDL') + \FDL\left(\xDL(\tDL\,)\right) + \noiseDL(\tDL),
\end{equation}
\end{widetext}
where $\FDL(\xDL) := \beta L \partial_x U(L \xDL) \equiv  \beta L  U'(L \xDL)$ and
$\noiseDL(\tDL) := \beta L\, \noise(\td  \tDL)$.
Using Eq.~\eqref{eq:DoubleExp_Dimensionless_Derivation_FDT}, it follows that the
autocorrelation of the dimensionless random force is given by
\begin{equation}
	\label{eq:DoubleExp_supp_dimlessFDT}
	\left\langle \noiseDL(\tDL) \noiseDL(\tDL')\right\rangle = 
		\sum_{i=1}^{N} \frac{\gamma_i}{\gamma} \frac{\td}{\tgi}
		\exp\left[-\frac{\td}{\tgi}| \tDL - \tDL'| \right].
\end{equation}
For $N=1$ we recover the case of  single-exponential memory, 
for $N=2$ and $\gamma_1 = \gamma_2$ we obtain the symmetric bi-exponential system that we  considered in the simulations.
For given potential $U$, Eqs.~\eqref{eq:DoubleExp_Dimensionless_Langevin_LangevinMultiExpDL}, \eqref{eq:DoubleExp_supp_dimlessFDT},
are fully determined by the dimensionless ratios $\tm/\td$, $\tgi/\td$, $\gamma_i/\gamma$.

\section{Transforming the GLE into a coupled system of Markovian equations}
\label{app:TransformingToMarkovian}

We now show that the dimensionless GLE with multi-exponential memory, Eqs.~\eqref{eq:DoubleExp_Dimensionless_Langevin_LangevinMultiExpDL}, 
\eqref{eq:DoubleExp_supp_dimlessFDT},
is equivalent to the Markovian coupled system of equations
\begin{align}
	\label{eq:MarkovianSystemA}
	m \ddot{x} &= -\partial_x V,\\
	\label{eq:MarkovianSystemB}
	0 &= -\gamma_i \dot{y}_i - \partial_i V + \xi_i, \qquad i\in \{1,...,N\},
\end{align}
where the $\xi_i$ are Markovian Gaussian random fields  with zero mean and variance
\begin{equation}
	\label{eq:DoubleExp_NumericalAlgorithmVariance}
	\beta \left\langle \xi_i(t) \xi_j(t) \right\rangle = 2 \gamma_i \delta_{i,j} \delta(t-t'), \qquad  i, j\in \{1,...,N\},
\end{equation}
the potential $V$ is given by
\begin{equation}
	\label{eq:MultidimPotential}
	V(x,y_1,...,y_N) = U(x) + \sum_{j=1}^N \frac{k_j}{2}(y_j - x)^2,
\end{equation}
and $\partial_i V$ denotes the partial derivative of $V$ w.r.t.~$y_i$.

Defining dimensionless positions $\xDL(\tDL) := x(\td  \tDL)/L$, $\yDL_i(\tDL) := y_i(\td  \tDL)/L$, inserting the potential 
Eq.~\eqref{eq:MultidimPotential} in Eqs.~\eqref{eq:MarkovianSystemA}, \eqref{eq:MarkovianSystemB}, introducing 
a rescaled time variable $t = \tDL \tau_{D}$, and multiplying by $\beta L$, we obtain
\begin{align}
	\label{eq:MarkovianSystemA_DL}
	\frac{\tm}{\td} \ddotxDL(\tDL\,) &= \FDL\left(\xDL(\tDL\,)\right) + \sum_{i=1}^N \frac{\td}{\tgi}\frac{\gamma_i}{\gamma} \left[\yDL_i(\tDL) - \xDL(\tDL\,)\right],\\
	\label{eq:MarkovianSystemB_DL}
	\dotyDL_i(\tDL\,) &= - \frac{\td}{\tgi} \yDL_i(\tDL\,) +  \frac{\td}{\tgi}\xDL(\tDL\,) + \sqrt{\frac{\gamma}{\gamma_i}}\xiDL_i(\tDL\,)
\quad 1 \leq i \leq N,
\end{align}
where $\FDL(\xDL) :=- \beta L \partial_x U(L \xDL) \equiv \beta L U'(L \xDL)$, $\tgi := \gamma_i/k_i$, and the 
correlators of the dimensionless random forces 
$\xiDL(\tDL) := \beta L\,  \sqrt{\gamma/\gamma_i} \xi(\td  \tDL)$ are given by
\begin{equation}
	\label{eq:MarkovianSystemC_DL}
	\left\langle\xiDL_i(\tDL\,)\xiDL_j(\tDL\,) \right\rangle = 2\delta_{i,j}\delta(\tDL-\tDL') \qquad i,j \in\{1,...,N\}.
\end{equation}

To show the equivalence of Eqs.~\eqref{eq:MarkovianSystemA_DL}, 
\eqref{eq:MarkovianSystemB_DL}, \eqref{eq:MarkovianSystemC_DL} to the GLE Eq.~\eqref{eq:DoubleExp_Dimensionless_Langevin_LangevinMultiExpDL}, \eqref{eq:DoubleExp_supp_dimlessFDT},
 we first note that
the solution to Eq.~\eqref{eq:MarkovianSystemB_DL} is given by
\begin{align}
\nonumber
	\yDL_i(\tDL\,) &= \yDL_i(0) \exp\left(-\frac{\td}{\tgi}\tDL\right) \\
&\qquad
		+ \frac{\td}{\tgi} \int_0^{\tDL}\mathrm{d}\tDL' 
			\exp\left[-\frac{\td}{\tgi}(\tDL-\tDL')\right] \xDL(\tDL')
\\ & \qquad
\nonumber
		+\sqrt{ \frac{\gamma}{\gamma_i}} \int_0^{\tDL}\mathrm{d}\tDL' 
			\exp\left[-\frac{\td}{\tgi}(\tDL-\tDL')\right] \xiDL_i(\tDL').
\end{align}
Using integration by parts on the second term of this expression and 
inserting the result into Eq.~\eqref{eq:MarkovianSystemA_DL}, we obtain
\begin{widetext}
\begin{align}
	\label{eq:MarkovianSystemDerivationA}
	\frac{\tm}{\td} \ddotxDL(\tDL) &=  -\sum_{i=1}^N \frac{\gamma_i}{\gamma} \frac{\td}{\tgi} \int_0^{\tDL}\mathrm{d}\tDL' \exp\left[-\frac{\td}{\tgi}\left( \tDL-\tDL'\right) \right] \dotxDL(\tDL') + \FDL(\xDL(\tDL)) 
	+ \noiseDL_R(\tDL),
\end{align}
\end{widetext}
where we define
\begin{align}
	\noiseDL_R(\tDL) 
	&=  
	\sum_{i=1}^N 	\left\{ \frac{\gamma_i}{\gamma} \frac{\td}{\tgi} \left[ \yDL_i(0)-\xDL(0)\right] \exp\left[-\frac{\td}{\tgi}\left( \tDL\,\right) \right] 
	\vphantom{+\sqrt{\frac{\gamma_i}{\gamma}} \frac{\td}{\tgi} \int_0^{\tDL}\mathrm{d}\tDL' 
			\exp\left[-\frac{\td}{\tgi}(\tDL-\tDL')\right] \xiDL_i(\tDL')
}
	\right.\\
&\qquad
\nonumber
\left.		+ \sqrt{\frac{\gamma_i}{\gamma}} 
\frac{\td}{\tgi} \int_0^{\tDL}\mathrm{d}\tDL' 
			\exp\left[-\frac{\td}{\tgi}(\tDL-\tDL')\right] \xiDL_i(\tDL')
			\right\}.
\end{align}
To obtain the equivalence of Eq.~\eqref{eq:MarkovianSystemDerivationA}
 to Eq.~\eqref{eq:DoubleExp_Dimensionless_Langevin_LangevinMultiExpDL},
  $\noiseDL_R$ needs to be a stochastic processes identical to $\noiseDL$.
For this, we assume that, for given $x(0)$, the $y_i(0)$ are distributed according to a Boltzmann
distribution $\rho(y_1(0),...,y_N(0)\,|\,x(0) ) \sim \exp\left[ -\beta V(x(0), y_1(0),...,y_N(0))\right]$, which using Eq.~\eqref{eq:MultidimPotential}
leads to  normally distributed $\yDL_i(0) -\xDL(0)$ with zero mean and covariance matrix
\begin{equation}
	\left\langle \left[ \yDL_i(0)-\xDL(0)\right]\left[\yDL_j(0)-\xDL(0)\right]\right\rangle = \delta_{i,j} \left(\frac{\gamma_i}{\gamma}\frac{\td}{\tgi}\right)^{-1}.
\end{equation}
With this, the mean of $\noiseDL_R$ is easily seen to be zero, and the variance is given by
\begin{widetext}
\begin{align}
	\left\langle \noiseDL_R(\tDL)\noiseDL_R(\tDL') \right\rangle &= 
		\sum_{i=1}^N  \left(\frac{\gamma_i}{\gamma}\frac{\td}{\tgi}\right)^2\exp\left[-\frac{\td}{\tgi} (\tDL + \tDL') \right] \left\langle \left( \yDL_i(0)-\xDL(0)\right)^2 \right\rangle\\
	&\quad \nonumber
	+ \sum_{i=1}^N 
	\frac{\gamma_i}{\gamma}\left(\frac{\td}{\tgi}\right)^2 
		\int_0^{\tDL}\mathrm{d}u\int_0^{\tDL'}\mathrm{d}u'\,
	\exp\left[-\frac{\td}{\tgi}(\tDL+\tDL'-u-u')\right]
		\langle \xiDL_i(u)\xiDL_i(u') \rangle \\
	&= \sum_{i=1}^N\exp\left[-\frac{\td}{\tgi} (\tDL + \tDL') \right] \frac{\gamma_i}{\gamma}\frac{\td}{\tgi} \nonumber
	\\&\quad
	+  \sum_{i=1}^N
	2\frac{\gamma_i}{\gamma}\left(\frac{\td}{\tgi}\right)^2 
		\int_0^{\min\{\tDL,\tDL'\}}\mathrm{d}u\,
	\exp\left[-\frac{\td}{\tgi}(\tDL+\tDL'-2u)\right] \\
	&= \sum_{i=1}^N\exp\left[-\frac{\td}{\tgi} (\tDL + \tDL') \right] \frac{\gamma_i}{\gamma}\frac{\td}{\tgi} \nonumber
	\\ & \quad
	+  \sum_{i=1}^N
	\frac{\gamma_i}{\gamma}\frac{\td}{\tgi} 
	\left\{ \exp\left[-\frac{\td}{\tgi}\left(\tDL+\tDL'-2\min\{\tDL,\tDL'\} \right)\right] -
		\exp\left[-\frac{\td}{\tgi}\left(\tDL+\tDL'\right)\right] \right\} \\
	&= \sum_{i=1}^N
	\frac{\gamma_i}{\gamma}\frac{\td}{\tgi} 
	 \exp\left[-\frac{\td}{\tgi}|\tDL-\tDL'|\right],
\end{align}
\end{widetext}
where we use that $\tDL+\tDL'-2\min\{\tDL,\tDL'\} = |\tDL - \tDL'|$.
Thus, $\noiseDL_R$ is a Gaussian stochastic process with the first two moments identical to those of the Gaussian
stochastic process $\noiseDL$,
so that $\noiseDL_R \equiv \noiseDL$.
For $N=2$ and $\gamma_1 = \gamma_2$, we obtain the system considered in the simulations in the main text.

To simulate Eqs.~\eqref{eq:MarkovianSystemA_DL}, \eqref{eq:MarkovianSystemB_DL} using
a Runge-Kutta scheme,  we introduce a further auxiliary variable $z$ to rewrite the equations as a system of first-order equations
\begin{align}
	\dotxDL(\tDL\,) &= z(\tDL\,),\\
	\frac{\tm}{\td} \dot{z}(\tDL\,) &= \sum_{i=1}^N \frac{\td}{\tgi}\frac{\gamma_i}{\gamma} \left[\yDL_j(\tDL) - \xDL(\tDL\,)\right] + \FDL\left(\xDL(\tDL\,)\right),\\
	\dotyDL_i(\tDL\,) &= - \frac{\td}{\tgi} \left[ \yDL_i(\tDL\,) -\xDL(\tDL\,)\right] + \sqrt{\frac{\gamma}{\gamma_i}}\xiDL_i(\tDL\,)
\quad 1 \leq i \leq N,
\end{align}

Note that  the equivalence of Eqs.~\eqref{eq:MarkovianSystemA}, \eqref{eq:MarkovianSystemB}, \eqref{eq:DoubleExp_NumericalAlgorithmVariance}, \eqref{eq:MultidimPotential}
 to Eqs.~\eqref{eq:DoubleExp_Dimensionless_Langevin_LangevinMultiExpDL}, \eqref{eq:DoubleExp_supp_dimlessFDT}, 
 physically means that multi-exponential memory can be interpreted as the result of the interaction of a particle $x$
  with $N$ independent reservoirs $y_i$ with finite relaxation times $\tgi$.

\section{Analysis of autocorrelation function}
\label{sec:DoubleExponential_PropagatorAnalysis}

In Fig.~\ref{fig:DoubleExp_Trajectories} we 
observe similarities
 between the dynamics in the asymmetric bi-exponential scenario $\tgOne \ll \tgTwo$ and the
  single-exponential scenario characterized by  the smaller memory time $\tgOne$,
   indicating that the smaller memory time dominates the  barrier-crossing kinetics.
 This can be rationalized by an asymptotic 
  analysis of the autocorrelation function  $C(t) \equiv \langle x(t)x(0) \rangle$ that characterizes
the particle motion within one potential well.
The following calculation is a generalization of the  analysis carried out for
the single-exponential memory kernel in Ref.~\cite{kappler_memory-induced_2018}.
We consider the GLE \eqref{eq:DoubleExp_LangevinDoubleExp} in a harmonic potential 
$U(x) \simeq K x^2/2$ and for times $t \gg \tgOne, \tgTwo$, so that we can replace the 
upper limit in the memory integral by infinity.
Fourier transforming Eq.~\eqref{eq:DoubleExp_LangevinDoubleExp}  and solving for $\FT{x}(\omega)$ yields
\begin{align}
	\label{eq:DoubleExp_FTSolution}
	\FT{x}(\omega) &= \frac{\FT{\noise}(\omega)}{K - m \omega^2 + i\omega \FT{\Gamma}_+(\omega)} \equiv \FT{Q}(\omega)\FT{\noise}(\omega),
\end{align}
where the half-sided Fourier transform $\FT{\Gamma}_+$ of the bi-exponential memory kernel $\Gamma(t)$ is given by
\begin{align}
	\FT{\Gamma}_+(\omega) &= \int_0^{\infty} \mathrm{d}t ~e^{-i\omega t}\Gamma(t) = \frac{\gamma}{2} \sum_{j=1}^2 \frac{1}{1+i\omega \tgj},
\end{align}
while  the full Fourier transform is
\begin{align}
	\FT{\Gamma}(\omega) &= \FT{\Gamma}_+(\omega) + \FT{\Gamma}_+(-\omega) = 
	\gamma \sum_{j=1}^2 \frac{1}{1+\omega^2 \tgj^2}.
\end{align}

Using Eq.~\eqref{eq:DoubleExp_FTSolution}, we calculate the autocorrelation function
 $C(t) \equiv \langle x(t)x(0)\rangle$ as
\begin{align}
	C(t) &\equiv \langle x(t)x(0) \rangle %\\
	%&
	= \int \frac{\mathrm{d}\omega}{2\pi} e^{i\omega t}~\int \frac{\mathrm{d}\omega'}{2\pi}~\langle \FT{x}(\omega) \FT{x}(\omega')\rangle \\
	&
	= \int \frac{\mathrm{d}\omega}{2\pi} e^{i\omega t}~\int \frac{\mathrm{d}\omega'}{2\pi}~ \FT{Q}(\omega) \FT{Q}(\omega') \langle \FT{\noise}(\omega) \FT{\noise}(\omega')\rangle \\
	&
= \beta^{-1} \int \frac{\mathrm{d}\omega}{2\pi} e^{i\omega t}~\int \frac{\mathrm{d}\omega'}{2\pi}~2 \pi \delta(\omega + \omega') \FT{\Gamma}(\omega) \FT{Q}(\omega) \FT{Q}(\omega')  \\
	&
= \beta^{-1} \int \frac{\mathrm{d}\omega}{2\pi} e^{i\omega t}~ \FT{\Gamma}(\omega) \FT{Q}(\omega) \FT{Q}(-\omega),
\end{align}
where we used that the Fourier transform of the generalized FDT is $\langle \FT{\noise}(\omega)\FT{\noise}(\omega')\rangle = \beta^{-1} ~2 \pi \delta(\omega + \omega') ~\FT{\Gamma}(\omega)$.
Thus the  Fourier transform of $C(t)$ is finally given by
\begin{widetext}
\begin{align}
	\beta \FT{C}(\omega) &= \FT{\Gamma}(\omega) \FT{Q}(\omega) \FT{Q}(-\omega) \\
	&=  \gamma \sum_{j=1}^2 \left[ \frac{1}{1+\omega^2 \tgj^2}  \nonumber
		\frac{1}{K- m \omega^2 + i\omega \gamma/2 \sum_{k=1}^2 (1+i \omega \tgk)^{-1} } \right. \\
		&\hspace*{3cm}
\left.	\times
		\frac{1}{K- m \omega^2 - i\omega \gamma/2 \sum_{l=1}^2 (1-i \omega \tgl)^{-1} } \right]\\
			\label{eq:DoubleExp_FTCorrelationFunction}
	&=  \gamma
		\sum_{j=1}^2 \left\{ (1+ \omega^2 \tgj^2)
		\left[
			 	(K-m \omega^2)^2
%	&\qquad 
				- \omega \gamma \sum_{k=1}^2 \omega \tgk(1+\omega^2\tgk^2)^{-1} \right. \right.\\
&\hspace*{4cm}
\left.				\left.+\frac{\omega^2 \gamma^2}{4} \sum_{k,l=1}^2 (1+i \omega \tgk)^{-1} (1-i \omega \tgl)^{-1}\right]\right\}^{-1}.
	\nonumber
\end{align}
\end{widetext}
For $\tgOne$, $\tgTwo$ large, we rewrite this as
\begin{align}
	\beta {\FT{C}(\omega)} &= \frac{\gamma}{\omega^2 \tgOne^2} \left[(K-m\omega^2)^2 + \sum_{j=1}^{2} \mathcal{O}\left( (\omega \tgj)^{-1} \right)\right]^{-1} 		
	\\&\hspace*{1cm}
\nonumber
	+\frac{\gamma}{\omega^2 \tgTwo^2} \left[(K-m\omega^2)^2 + \sum_{j=1}^{2} \mathcal{O}\left( (\omega \tgj)^{-1} \right)\right]^{-1},
\end{align}
so that to order $(\omega \tgi)^2$ the autocorrelation function  is just the sum of
two terms corresponding to the two single-exponential memory contributions. 
For $\tgOne \ll \tgTwo$, the term involving $\tgOne$ dominates and we obtain
\begin{align}
	\label{eq:DoubleExp_DoubleExpPropFinal}
	\beta {\FT{C}(\omega)} &\approx \frac{\gamma}{\omega^2 \tgOne^2} \left[(K-m\omega^2)^2 \right]^{-1}.
\end{align}
The result is independent of $\tgTwo$, showing that the dynamics is indeed dominated by $\tgOne$, which
explains the behavior of the trajectories shown in Fig.~\ref{fig:DoubleExp_Trajectories} (c), (f).

\section{Multi-exponential mean first-passage time in the energy diffusion regime}
\label{app:inverse_additivity}

In the energy-diffusion limit and for high barriers, the barrier-crossing rate is given by \cite{talkner_transition_1988}
\begin{equation}
	\label{eq:TalknerRate}
	k_{\mathrm{ED}} = \frac{\beta \omega_0}{8\pi}e^{-\beta U_0} \int_{-\infty}^{\infty}\mathrm{d}t \int_{-\infty}^{\infty} \mathrm{d}s~
	\Gamma(|t-s|) \dot{x}_{0}(t) \dot{x}_{0}(s),
\end{equation}
where $\omega_0 = \sqrt{U''(-L)/m}$ is the well frequency, and $\dot{x}_{0}$ is the velocity of the particle of the undamped system
described by Eq.~\eqref{eq:DoubleExp_LangevinDoubleExp} without memory kernel $\Gamma$ and random force $\noise$,
starting at the barrier top with $\dot{x}_0(0) = 0^{-}$ and traversing the left well once \cite{talkner_transition_1988}.
Since Eq.~\eqref{eq:TalknerRate} is linear in $\Gamma$, for a multi-exponential kernel
 it immediately follows that
\begin{equation}
	\label{eq:EDsum_appendix}
	k_{\mathrm{ED}} = \sum_{i} k_{\mathrm{ED}}^{(i)},
\end{equation}
 where $k_{\mathrm{ED}}^{(i)}$ is the single-exponential barrier-crossing rate corresponding the $i$-th exponential.
From this and the relation $\tf \sim 1/k$ between MFPT $\tf$ and barrier-crossing 
rate $k$ \cite{reimann_universal_1999,kappler_memory-induced_2018}, 
the second sum in Eq.~\eqref{eq:tf_multi} follows.

\section{Comparison of numerical MFPTs to GH theory}
\label{app:GH}

\begin{figure*}[ht]
	\includegraphics[width=0.30\textwidth]{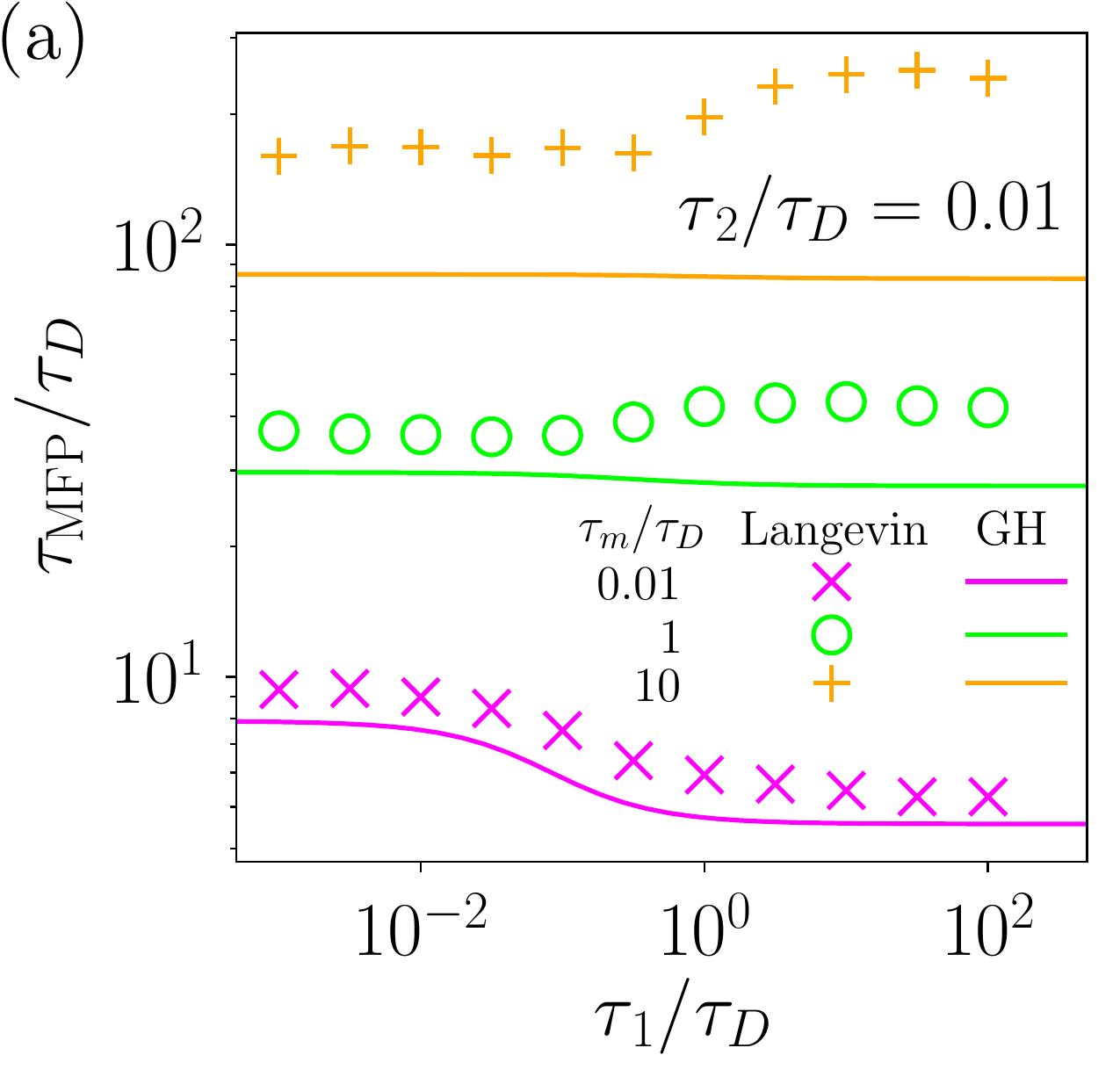}
	~~~~
	\includegraphics[width=0.30\textwidth]{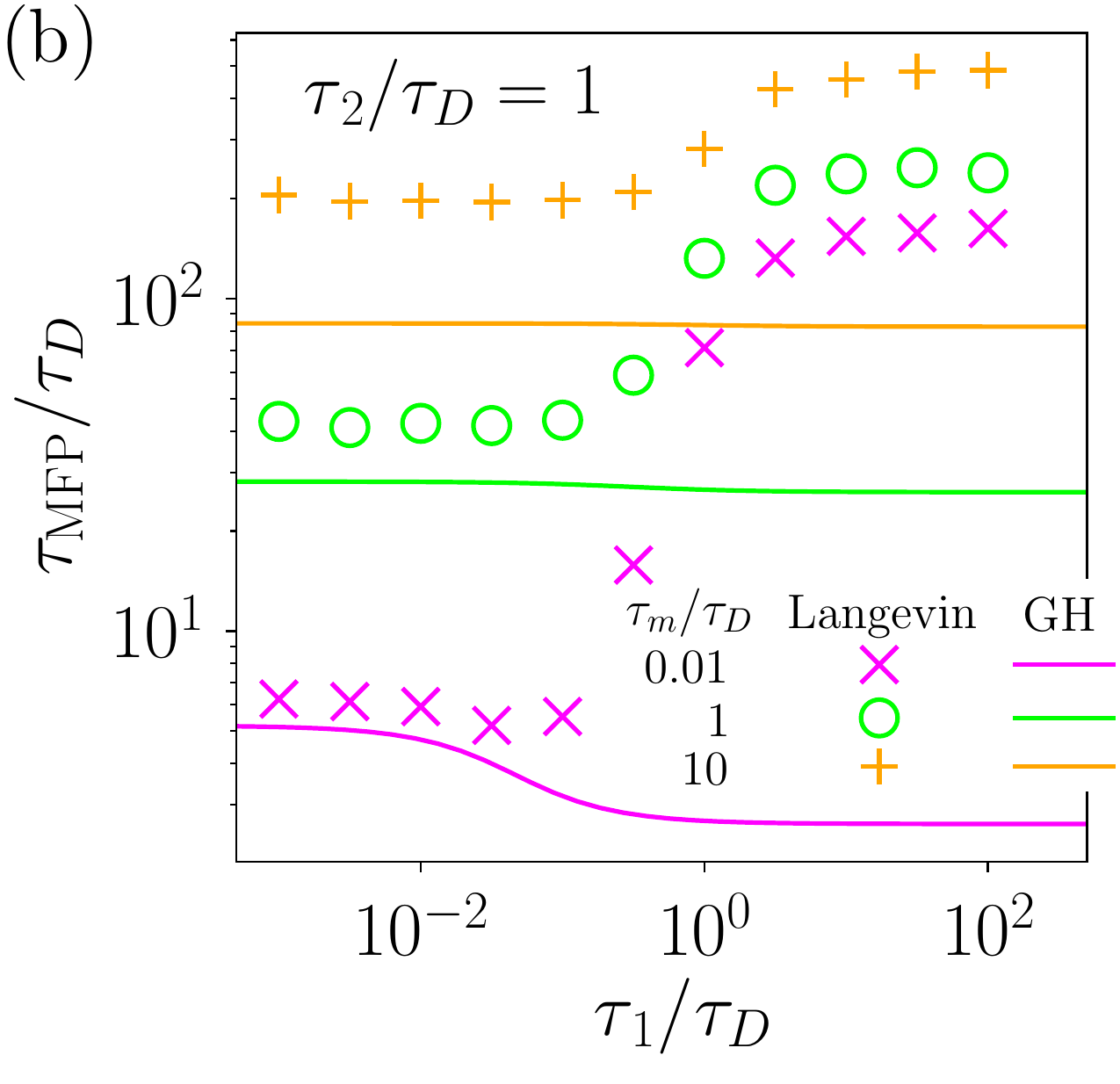}
	~~~~
	\includegraphics[width=0.30\textwidth]{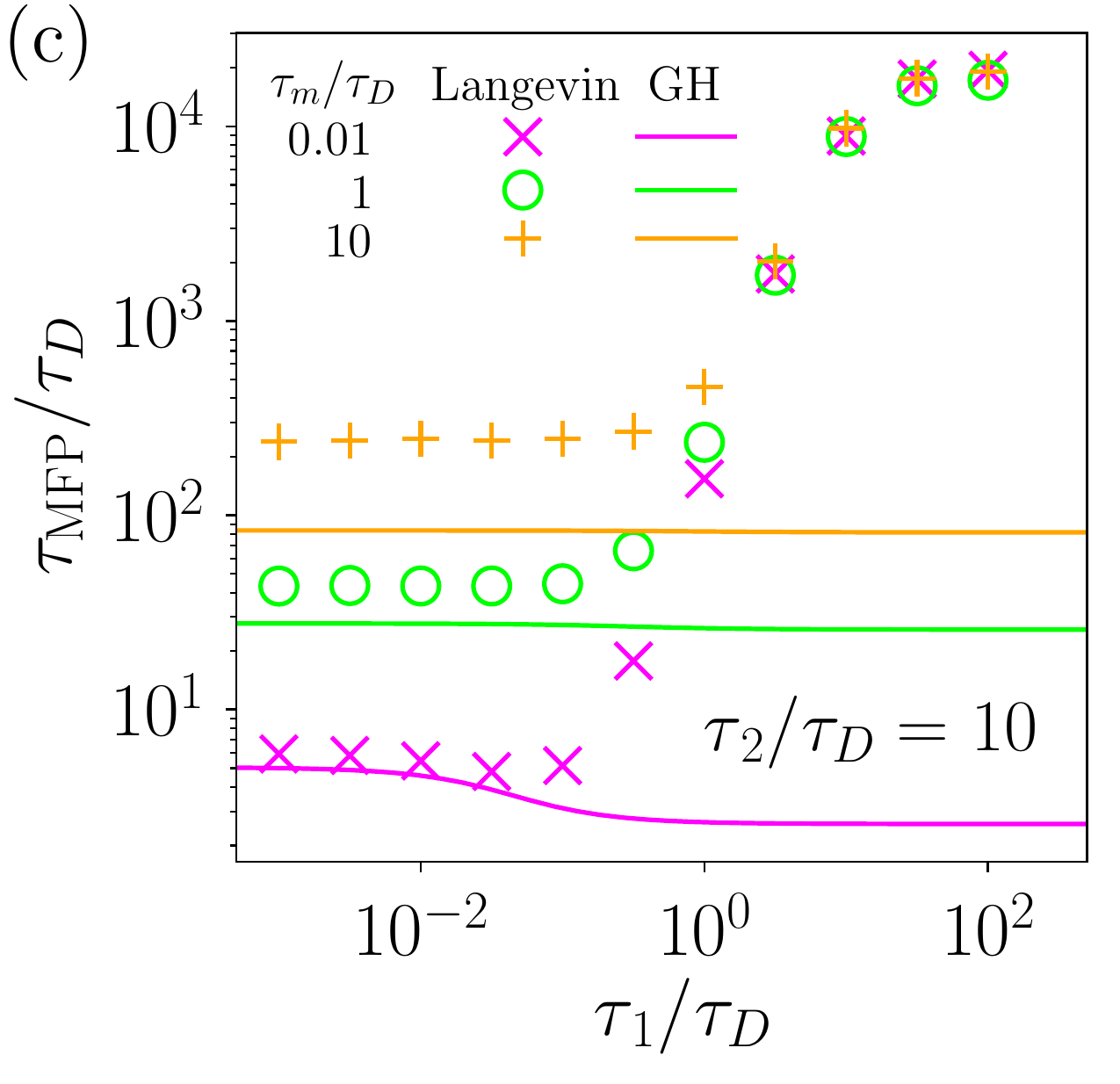}
\par
\vspace{0.4cm}
	\includegraphics[width=0.30\textwidth]{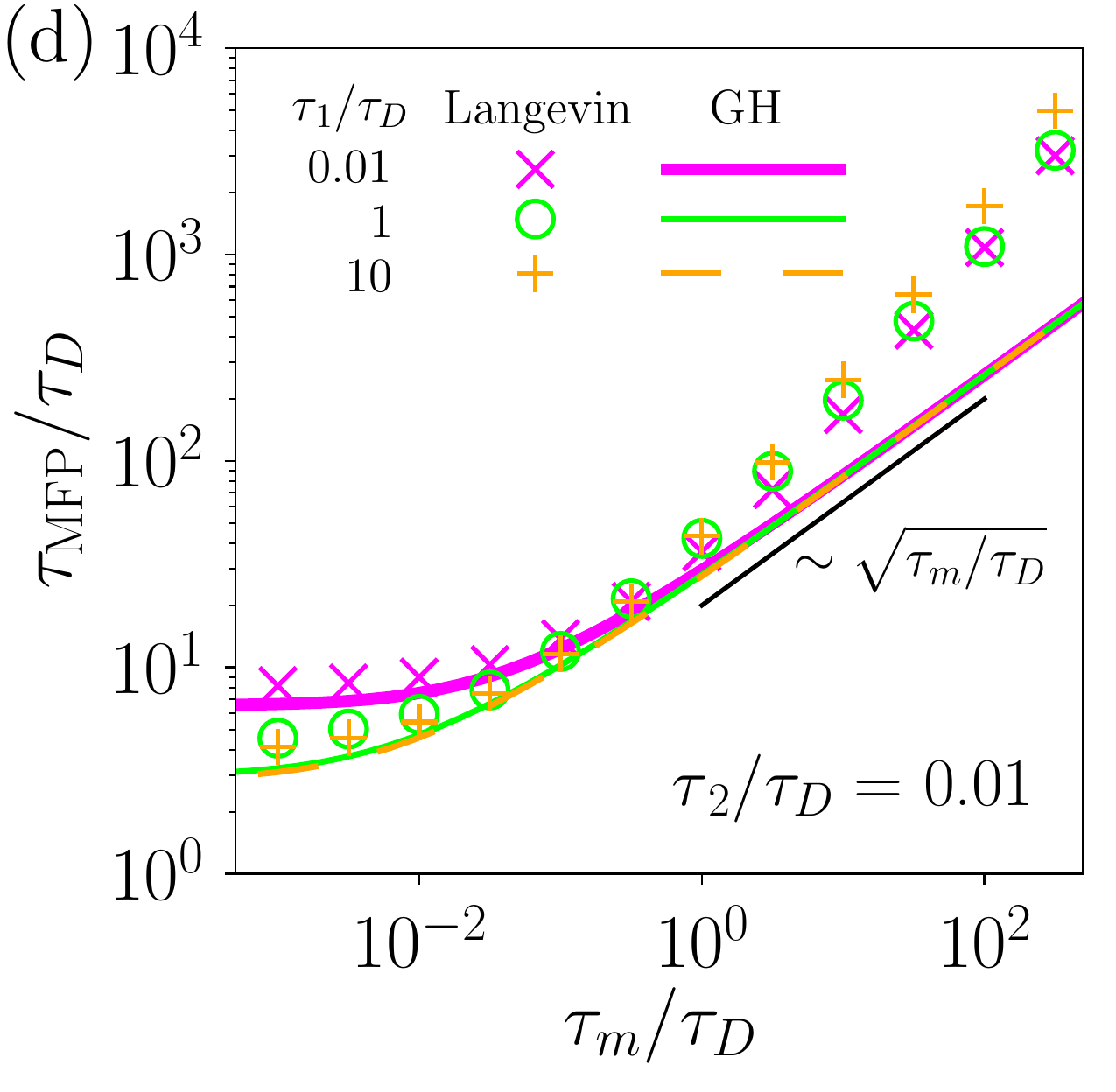}
	~~~~
	\includegraphics[width=0.30\textwidth]{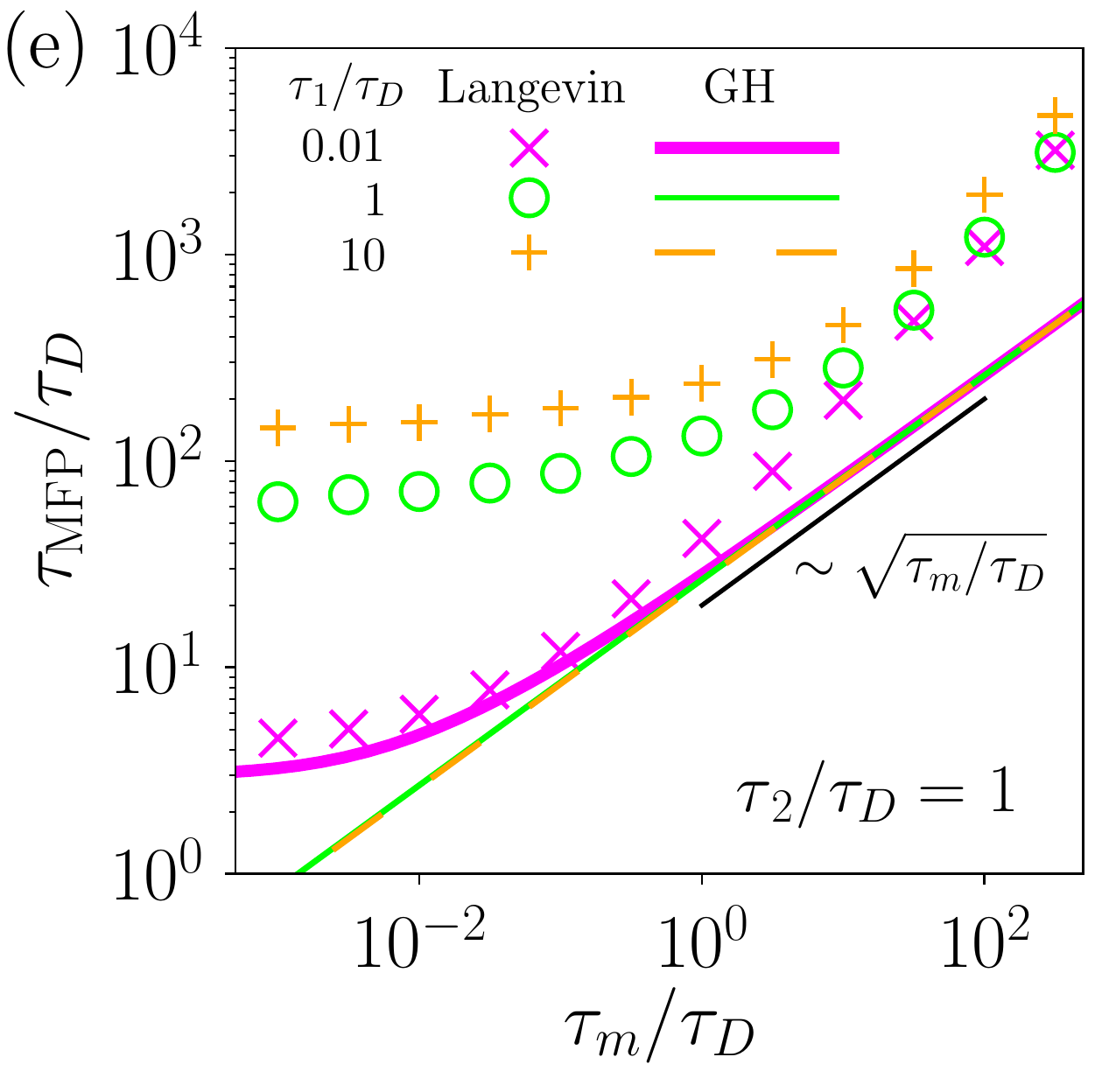}
	~~~~
	\includegraphics[width=0.30\textwidth]{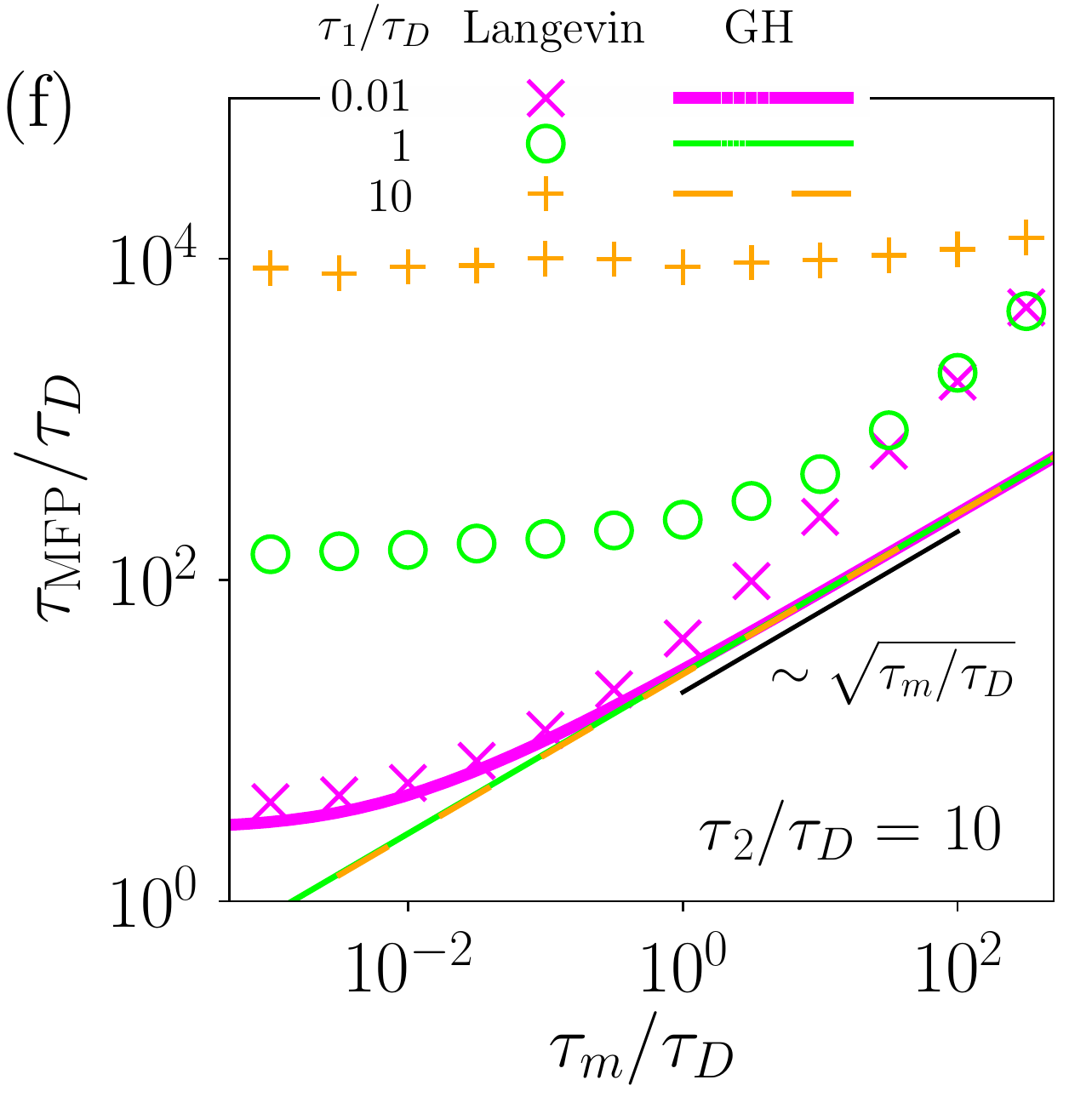}
\caption{
{Comparison of numerical MFPTs to Grote-Hynes (GH) theory.}
{(a)-(c)} Colored symbols denote the rescaled MFPT $\tf/\td$ as function of $\tgOne/\td$ for several values of $\tgTwo/\td$
and fixed $\tm/\td$, given by (a) $\tm/\td = 0.01$, (b) $\tm/\td= 1$, (c) $\tm/\td = 10$.
The numerical data is the same as shown in Fig.~\ref{fig:ComparisonTg2}. 
The colored lines represent  GH theory, evaluated for a bi-exponential memory kernel using the simulation parameters.
{(d)-(f)} Colored symbols denote the rescaled MFPT $\tf/\td$ as function of $\tm/\td$ for several values of $\tgTwo/\td$
and fixed $\tgOne/\td$, given by (a) $\tgOne/\td = 0.01$, (b) $\tgOne/\td= 1$, (c) $\tgOne/\td = 10$.
The numerical data is a replot of the data shown in Fig.~\ref{fig:ComparisonTg1}.
The colored lines represent  GH theory, evaluated for a bi-exponential memory kernel using the 
simulation  parameters.
The black lines indicate the transition-state-theory (TST) scaling $\tf/\td \sim \sqrt{\tm/\td}$ which  GH theory attains 
in the energy-diffusion regime \cite{grote_stable_1980}.
All data is obtained using $\beta U_0 = 3$.
}
\label{fig:DoubleExp_GH_comparison}
\end{figure*}

In Fig.~\ref{fig:DoubleExp_GH_comparison} we compare numerical MFPTs to Grote-Hynes (GH) theory \cite{grote_stable_1980}.
Consistent with previous results \cite{kappler_memory-induced_2018}, we find that GH theory only describes the numerical MFPTs
in the high-friction regime $\tm/\td \ll 1$ and if both memory times are small, $\tgOne/\td, \tgTwo/\td \ll 1$.
If either $\tm/\td \gg 1$ or $\min\{\tgOne/\td, \tgTwo/\td\} \gg 1$, then GH theory 
reduces to the transition state theory (TST) limit where $\tf/\td \sim \sqrt{\tm/\td}$,
 as can be seen in Figs.~\ref{fig:DoubleExp_GH_comparison} (d)-(f).
Thus, also for bi-exponential memory GH theory only yields accurate results in the high-friction short-memory regime.

 \textcolor{black}{The disagreement between our Langevin results and GH theory is not surprising, since
the latter is by construction not intended for the energy diffusion regime.
More specifically, GH theory considers a GLE close to the barrier top, 
so that the theory does not include memory effects from the motion of the particle in the potential well;
the GLE model we consider here has a global position-indepedent memory function \cite{pollak_theory_1989}.
Therefore, only in or close to the overdamped regime, 
where memory effects from motion in the well are not relevant to the dynamics
close to the barrier top, do the predictions of GH theory and our numerical results agree.
}

\section{Global comparison of numerical data to Eq.~\eqref{eq:tf_multi} \textcolor{black}{at $\beta U_0 = 3$}}
\label{app:GlobalComparison}

\begin{figure*}
\centering
	\includegraphics[width=\textwidth]{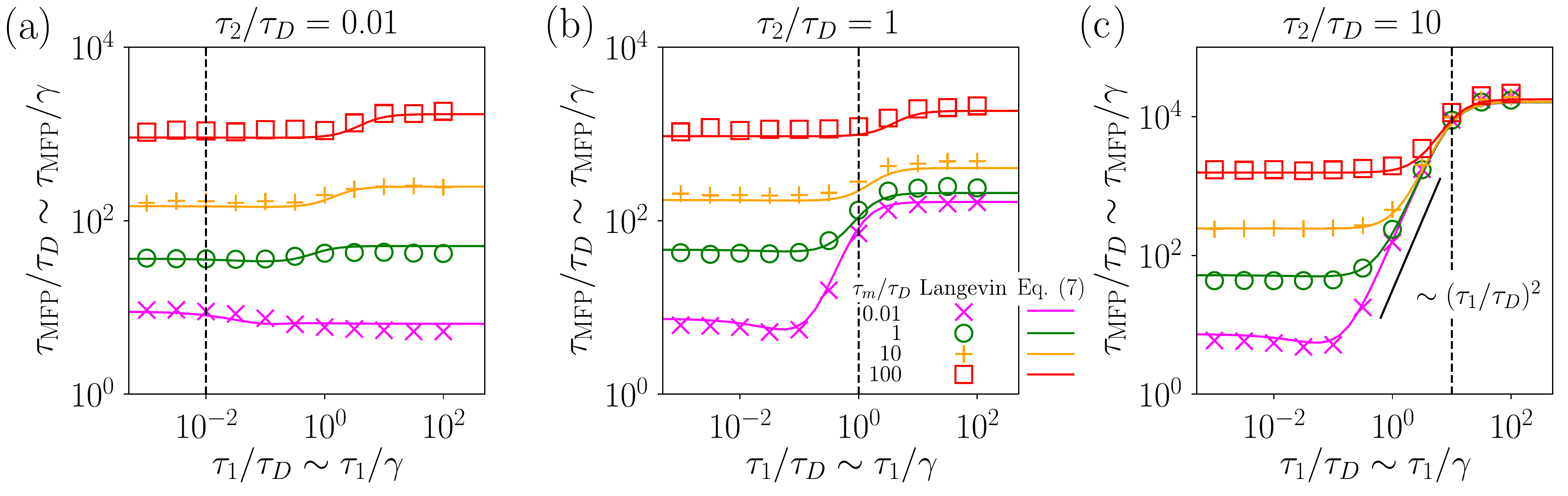}
\caption{
{Simulation results for the MFPT at fixed $\tgTwo/\td$.}
 Colored symbols denote the rescaled MFPT $\tf/\td$ as function of $\tgOne/\td$ for several values of $\tm/\td$
and fixed $\tgTwo/\td$, given by (a) $\tgTwo/\td = 0.01$, (b) $\tgTwo/\td= 1$, (c) $\tgTwo/\td = 10$.
The legend for all subplots is given in subplot (b).
The colored lines represent the crossover  formula Eq.~\eqref{eq:tf_multi},
the vertical dashed lines  denote where  $\tgOne = \tgTwo$.
The black bar in (c) indicates the scaling $\tf \sim \tgOne^2$ which corresponds to
 the single-exponential long-memory regime.
All data is obtained using $\beta U_0 = 3$.
}
\label{fig:ComparisonTg2}
\end{figure*}

\begin{figure*}
\centering
	\includegraphics[width=\textwidth]{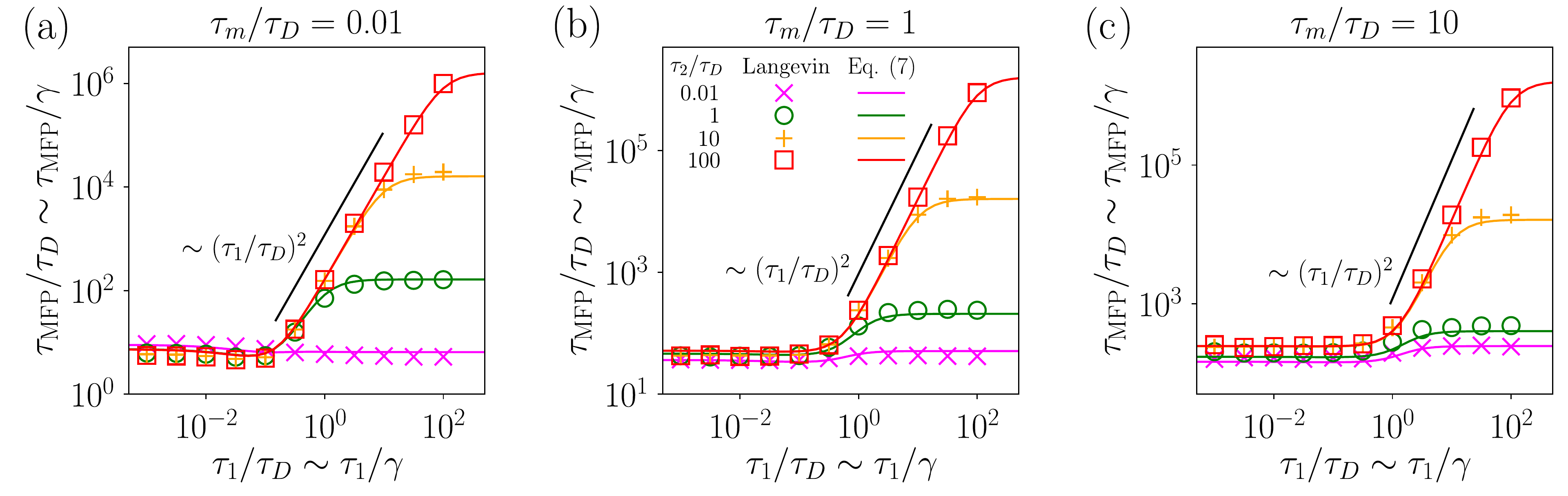}
\caption{
{Simulation results for the MFPT at fixed $\tm/\td$.}
 Colored symbols denote the rescaled MFPT $\tf/\td$ as function of $\tgOne/\td$ for several values of $\tgTwo/\td$
and fixed $\tm/\td$, given by (a) $\tm/\td = 0.01$, (b) $\tm/\td= 1$, (c) $\tm/\td = 10$.
The legend for all subplots is given in subplot (b).
The colored lines represent the crossover  formula Eq.~\eqref{eq:tf_multi},
the black bars indicate the scaling $\tf \sim \tgOne^2$, which corresponds to
 the single-exponential long-memory regime.
 All data is obtained using $\beta U_0 = 3$.
}
\label{fig:ComparisonTm}
\end{figure*}

\begin{figure*}
\centering
	\includegraphics[width=\textwidth]{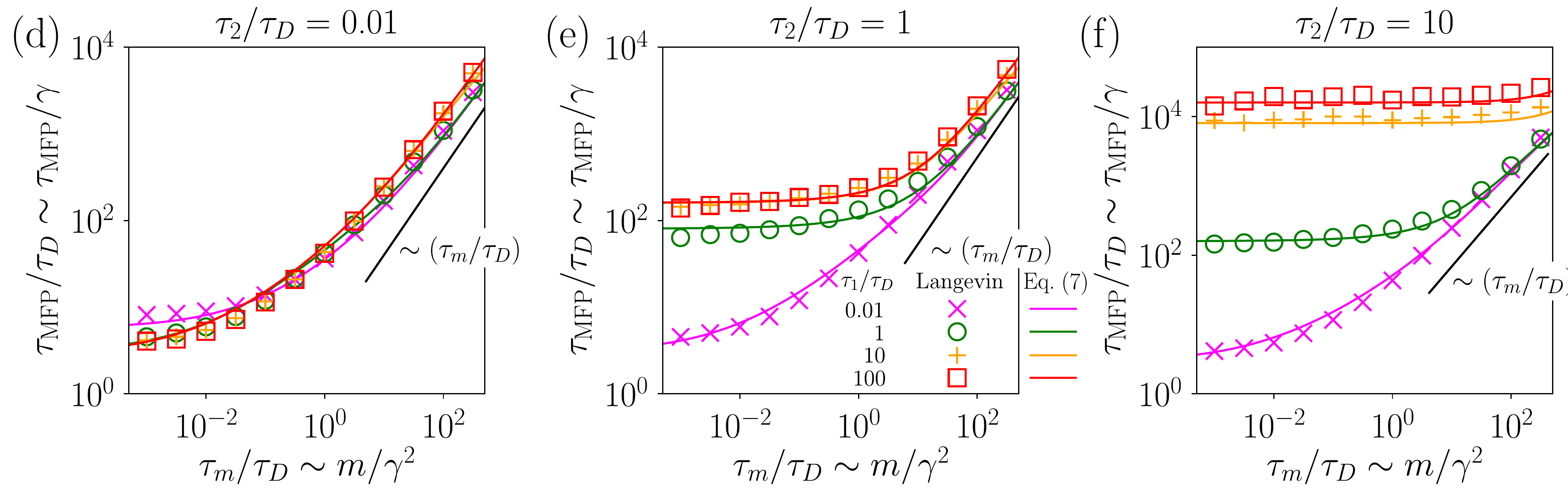}
\caption{
{Simulation results for the MFPT at fixed $\tgTwo/\td$.}
 Colored symbols denote the rescaled MFPT $\tf/\td$ as function of $\tm/\td$ for several values of $\tgOne/\td$
and fixed $\tgTwo/\td$, given by (a) $\tgTwo/\td = 0.01$, (b) $\tgTwo/\td= 1$, (c) $\tgTwo/\td = 10$.
The legend for all subplots is given in subplot (b).
The colored lines represent the crossover  formula Eq.~\eqref{eq:tf_multi},
the black bars indicate the scaling $\tf \sim \tm$ which corresponds to
 the Markovian inertial regime.
All data is obtained using $\beta U_0 = 3$.
}
\label{fig:ComparisonTg1}
\end{figure*}

In Figs.~\ref{fig:ComparisonTg2}, \ref{fig:ComparisonTm}, \ref{fig:ComparisonTg1} we compare predictions
 of the crossover  formula Eq.~\eqref{eq:tf_multi} for $\tf$ to numerical results.
 These figures, which we discuss in detail in the following paragraphs, show that 
 Eq.~\eqref{eq:tf_multi} quantitively describes the bi-exponential MFPTs
over the whole parameter range \textcolor{black}{of $\tm/\td$, $\tgOne/\td$, $\tgTwo/\td$} considered,
and in particular that the MFPT
is in the non-Markovian regime  dominated by the shorter memory time.

Figure \ref{fig:ComparisonTg2} shows $\tf/\td$ as function of $\tgOne/\td$ 
for fixed values of $\tm/\td$, $\tgTwo/\td$.
In the figure we see that for $\tgOne \ll \tgTwo$, 
i.e., to the left of the  vertical dashed lines that denote $\tgOne = \tgTwo$, 
the MFPT  behaves similar to the single-exponential MFPT shown in  Fig.~\ref{fig:MFPTs} (b).
In particular, for $\tm/\td = 0.01$ (magenta crosses) the memory acceleration regime can be 
seen in Fig.~\ref{fig:ComparisonTg2} (b), (c) for
$\tgOne/\td \approx 0.1$, followed by the power-law scaling $\tf/\td \sim (\tgOne/\td)^2$ 
for $0.1 \lesssim \tgOne/\td \lesssim \tgTwo/\td$.
As $\tgOne \gtrsim \tgTwo$, i.e., to the right of the  vertical dashed lines,
 the MFPT becomes independent of $\tgOne$. %, 
This can prominently be seen in Fig.~\ref{fig:ComparisonTg2} (a), where $\tgTwo/\td = 0.01$ is small and 
$\tf/\td$ is
 almost constant for $\tgOne > \tgTwo$. %
Throughout Fig.~\ref{fig:ComparisonTg2} the crossover formula 
Eq.~\eqref{eq:tf_multi} describes the numerical results
very well.

Figure \ref{fig:ComparisonTm} shows $\tf/\td$ as function of $\tgOne/\td$
for fixed values of $\tgTwo/\td$, $\tm/\td$.
In Fig.~\ref{fig:ComparisonTm} (a), we again see single-exponential behavior
reminiscent of Fig.~\ref{fig:MFPTs} (b).
More explicitly, for $\tgTwo/\td = 1$, $10$, $100$, we 
observe a slight dip in $\tf/\td$ 
for $\tgOne/\td \approx 0.1$, followed by power-law scaling $\tf/\td \sim (\tgOne/\td)^2$ for 
$0.1 \lesssim \tgOne/\td \lesssim \tgTwo/\td$.
As $\tgOne \gtrsim \tgTwo$, $\tf/\td$ saturates to a value
determined by $\tm/\td$, $\tgTwo/\td$.
For $\tgTwo/\td = 0.01$ (magenta crosses), Fig.~\ref{fig:ComparisonTm} (a) contains
no regime where $\tgOne \ll \tgTwo$,
and $\tf/\td$ is almost independent of $\tgOne/\td$ throughout.
Figs.~\ref{fig:ComparisonTm} (b), (c) show similar behavior, and as expected from the single-exponential
data for $\tm/\td = 1$, $10$ (see Fig.~\ref{fig:MFPTs} (a)), a dip in the MFPT for $\tgOne/\td \approx 0.1$
cannot be observed in the logarithmic representation of $\tf/\td$.
Again, the crossover  formula Eq.~\eqref{eq:tf_multi} describes the data very 
accurately throughout.

Figure \ref{fig:ComparisonTg1} shows $\tf/\td$ as a function of $\tm/\td$ for 
fixed values of $\tgOne/\td$, $\tgTwo/\td$.
In all three plots, the MFPT displays a similar $\tm/\td$-dependence as the single-exponential 
case, c.f.~Fig.~\ref{fig:MFPTs} (a).
For $\tgTwo/\td = 0.01$ we see in Fig.~\ref{fig:ComparisonTg1} (a) 
that $\tf/\td$ is almost independent of $\tgOne/\td$,
consistent with the picture that $\tf/\td$ is determined by the shorter memory time $\tgTwo$.
On the other hand, for large $\tgTwo/\td = 10$ the MFPT shown in Fig.~\ref{fig:ComparisonTg1} (c)
depends very much on $\tgOne/\td$ and is qualitatively identical to the single-exponential MFPT shown
in Fig.~\ref{fig:MFPTs} (a).
Also for this case, we see that the crossover  formula Eq.~\eqref{eq:tf_multi} remains accurate.

\section{\textcolor{black}{Comparison of numerical data to Eq.~\eqref{eq:tf_multi} for varying barrier height $\beta U_0$}}
\label{app:BarrierHeight}

\begin{figure*}
\centering
	\includegraphics[width=0.9\textwidth]{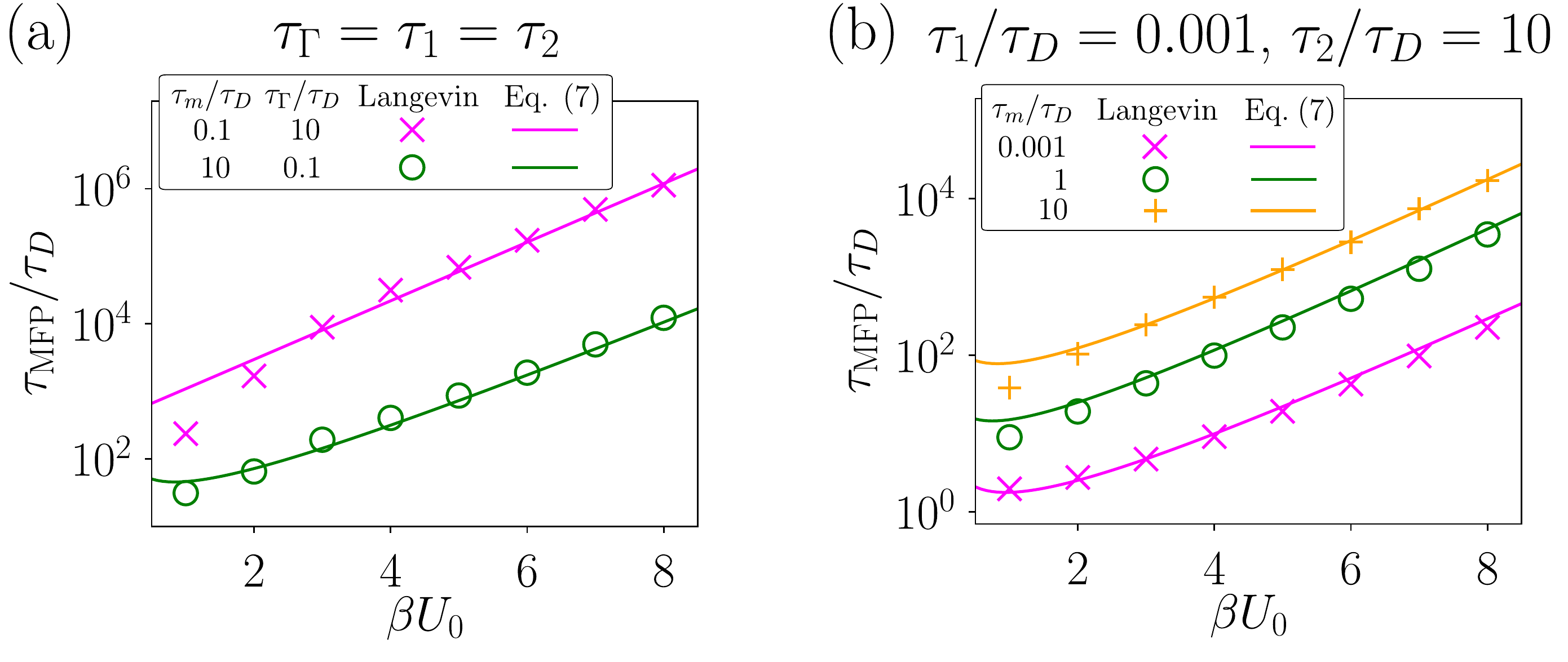}
\caption{
\textcolor{black}{
Simulation results for the MFPT as a function of barrier height $\beta U_0$.
 Colored symbols denote the rescaled MFPT $\tf/\td$, obtained from numerical simulations,
 as function of $\beta U_0$ for fixed values of $\tm/\td$, $\tgOne/\td$, $\tgTwo/\td$;
 colored solid lines represent the crossover  formula Eq.~\eqref{eq:tf_multi}.  
 (a) Symmetric scenario $\tg \equiv \tgOne = \tgTwo$. 
 (b) Asymmetric scenario with $\tgOne/\td = 0.001$, $\tgTwo/\td = 10$.
Note that the symmetric scenario has previously been considered
  in the literature \cite{kappler_memory-induced_2018}.
}
}
\label{fig:ComparisonU}
\end{figure*}

\textcolor{black}{
In Fig.~\ref{fig:ComparisonU} we compare predictions
 of the crossover  formula Eq.~\eqref{eq:tf_multi} for $\tf$ to numerical results,
 varying the barrier height $\beta U_0$.
While  Fig.~\ref{fig:ComparisonU}  (a) considers the symmetric case ($\tg \equiv \tgOne = \tgTwo$),
Fig.~\ref{fig:ComparisonU} (b) considers the asymmetric
scenario ($0.001 = \tgOne/\td \ll \tgTwo/\td = 10$).
The agreement between Eq.~\eqref{eq:tf_multi} and numerical data in Fig.~\ref{fig:ComparisonU} shows
that our heuristic formula
 accurately describes the Langevin data 
 for barrier heights larger than $\beta U_0 \approx 2$ throughout.
}

\begin{thebibliography}{54}%
\makeatletter
\providecommand \@ifxundefined [1]{%
 \@ifx{#1\undefined}
}%
\providecommand \@ifnum [1]{%
 \ifnum #1\expandafter \@firstoftwo
 \else \expandafter \@secondoftwo
 \fi
}%
\providecommand \@ifx [1]{%
 \ifx #1\expandafter \@firstoftwo
 \else \expandafter \@secondoftwo
 \fi
}%
\providecommand \natexlab [1]{#1}%
\providecommand \enquote  [1]{``#1''}%
\providecommand \bibnamefont  [1]{#1}%
\providecommand \bibfnamefont [1]{#1}%
\providecommand \citenamefont [1]{#1}%
\providecommand \href@noop [0]{\@secondoftwo}%
\providecommand \href [0]{\begingroup \@sanitize@url \@href}%
\providecommand \@href[1]{\@@startlink{#1}\@@href}%
\providecommand \@@href[1]{\endgroup#1\@@endlink}%
\providecommand \@sanitize@url [0]{\catcode `\\12\catcode `\$12\catcode
  `\&12\catcode `\#12\catcode `\^12\catcode `\_12\catcode `\%12\relax}%
\providecommand \@@startlink[1]{}%
\providecommand \@@endlink[0]{}%
\providecommand \url  [0]{\begingroup\@sanitize@url \@url }%
\providecommand \@url [1]{\endgroup\@href {#1}{\urlprefix }}%
\providecommand \urlprefix  [0]{URL }%
\providecommand \Eprint [0]{\href }%
\providecommand \doibase [0]{http://dx.doi.org/}%
\providecommand \selectlanguage [0]{\@gobble}%
\providecommand \bibinfo  [0]{\@secondoftwo}%
\providecommand \bibfield  [0]{\@secondoftwo}%
\providecommand \translation [1]{[#1]}%
\providecommand \BibitemOpen [0]{}%
\providecommand \bibitemStop [0]{}%
\providecommand \bibitemNoStop [0]{.\EOS\space}%
\providecommand \EOS [0]{\spacefactor3000\relax}%
\providecommand \BibitemShut  [1]{\csname bibitem#1\endcsname}%
\let\auto@bib@innerbib\@empty
%</preamble>
\bibitem [{\citenamefont {Kramers}(1940)}]{kramers_brownian_1940}%
  \BibitemOpen
  \bibfield  {author} {\bibinfo {author} {\bibfnamefont {H.~A.}\ \bibnamefont
  {Kramers}},\ }\href@noop {} {\bibfield  {journal} {\bibinfo  {journal}
  {Physica}\ }\textbf {\bibinfo {volume} {7}},\ \bibinfo {pages} {284}
  (\bibinfo {year} {1940})}\BibitemShut {NoStop}%
\bibitem [{\citenamefont {Chandler}(1986)}]{chandler1986}%
  \BibitemOpen
  \bibfield  {author} {\bibinfo {author} {\bibfnamefont {D.}~\bibnamefont
  {Chandler}},\ }\href@noop {} {\bibfield  {journal} {\bibinfo  {journal}
  {Journal of Statistical Physics}\ }\textbf {\bibinfo {volume} {42}},\
  \bibinfo {pages} {49} (\bibinfo {year} {1986})}\BibitemShut {NoStop}%
\bibitem [{\citenamefont {Berne}\ \emph {et~al.}(1988)\citenamefont {Berne},
  \citenamefont {Borkovec},\ and\ \citenamefont {Straub}}]{berne1988}%
  \BibitemOpen
  \bibfield  {author} {\bibinfo {author} {\bibfnamefont {B.~J.}\ \bibnamefont
  {Berne}}, \bibinfo {author} {\bibfnamefont {M.}~\bibnamefont {Borkovec}}, \
  and\ \bibinfo {author} {\bibfnamefont {J.~E.}\ \bibnamefont {Straub}},\
  }\href@noop {} {\bibfield  {journal} {\bibinfo  {journal} {Journal of
  Physical Chemistry}\ }\textbf {\bibinfo {volume} {92}},\ \bibinfo {pages}
  {3711} (\bibinfo {year} {1988})}\BibitemShut {NoStop}%
\bibitem [{\citenamefont {Hänggi}\ \emph {et~al.}(1990)\citenamefont
  {Hänggi}, \citenamefont {Talkner},\ and\ \citenamefont
  {Borkovec}}]{hanggi_reaction-rate_1990}%
  \BibitemOpen
  \bibfield  {author} {\bibinfo {author} {\bibfnamefont {P.}~\bibnamefont
  {Hänggi}}, \bibinfo {author} {\bibfnamefont {P.}~\bibnamefont {Talkner}}, \
  and\ \bibinfo {author} {\bibfnamefont {M.}~\bibnamefont {Borkovec}},\ }\href
  {\doibase 10.1103/RevModPhys.62.251} {\bibfield  {journal} {\bibinfo
  {journal} {Reviews of Modern Physics}\ }\textbf {\bibinfo {volume} {62}},\
  \bibinfo {pages} {251} (\bibinfo {year} {1990})}\BibitemShut {NoStop}%
\bibitem [{\citenamefont {Best}\ and\ \citenamefont
  {Hummer}(2006)}]{best_diffusive_2006}%
  \BibitemOpen
  \bibfield  {author} {\bibinfo {author} {\bibfnamefont {R.}~\bibnamefont
  {Best}}\ and\ \bibinfo {author} {\bibfnamefont {G.}~\bibnamefont {Hummer}},\
  }\href {\doibase 10.1103/PhysRevLett.96.228104} {\bibfield  {journal}
  {\bibinfo  {journal} {Physical Review Letters}\ }\textbf {\bibinfo {volume}
  {96}} (\bibinfo {year} {2006}),\ 10.1103/PhysRevLett.96.228104}\BibitemShut
  {NoStop}%
\bibitem [{\citenamefont {Zwanzig}(1961)}]{zwanzig_memory_1961}%
  \BibitemOpen
  \bibfield  {author} {\bibinfo {author} {\bibfnamefont {R.}~\bibnamefont
  {Zwanzig}},\ }\href {\doibase 10.1103/PhysRev.124.983} {\bibfield  {journal}
  {\bibinfo  {journal} {Physical Review}\ }\textbf {\bibinfo {volume} {124}},\
  \bibinfo {pages} {983} (\bibinfo {year} {1961})}\BibitemShut {NoStop}%
\bibitem [{\citenamefont {Mori}(1965)}]{mori_transport_1965}%
  \BibitemOpen
  \bibfield  {author} {\bibinfo {author} {\bibfnamefont {H.}~\bibnamefont
  {Mori}},\ }\href {\doibase 10.1143/PTP.33.423} {\bibfield  {journal}
  {\bibinfo  {journal} {Progress of Theoretical Physics}\ }\textbf {\bibinfo
  {volume} {33}},\ \bibinfo {pages} {423} (\bibinfo {year} {1965})}\BibitemShut
  {NoStop}%
\bibitem [{\citenamefont {Jung}\ \emph {et~al.}(2017)\citenamefont {Jung},
  \citenamefont {Hanke},\ and\ \citenamefont {Schmid}}]{Schmid_2017}%
  \BibitemOpen
  \bibfield  {author} {\bibinfo {author} {\bibfnamefont {G.}~\bibnamefont
  {Jung}}, \bibinfo {author} {\bibfnamefont {M.}~\bibnamefont {Hanke}}, \ and\
  \bibinfo {author} {\bibfnamefont {F.}~\bibnamefont {Schmid}},\ }\href
  {\doibase 10.1021/acs.jctc.7b00274} {\bibfield  {journal} {\bibinfo
  {journal} {Journal of Chemical Theory and Computation}\ }\textbf {\bibinfo
  {volume} {13}},\ \bibinfo {pages} {2481} (\bibinfo {year} {2017})},\ \bibinfo
  {note} {pMID: 28505440},\ \Eprint
  {http://arxiv.org/abs/https://doi.org/10.1021/acs.jctc.7b00274}
  {https://doi.org/10.1021/acs.jctc.7b00274} \BibitemShut {NoStop}%
\bibitem [{\citenamefont {Jung}\ \emph {et~al.}(2018)\citenamefont {Jung},
  \citenamefont {Hanke},\ and\ \citenamefont {Schmid}}]{Schmid_2018}%
  \BibitemOpen
  \bibfield  {author} {\bibinfo {author} {\bibfnamefont {G.}~\bibnamefont
  {Jung}}, \bibinfo {author} {\bibfnamefont {M.}~\bibnamefont {Hanke}}, \ and\
  \bibinfo {author} {\bibfnamefont {F.}~\bibnamefont {Schmid}},\ }\href
  {\doibase 10.1039/C8SM01817K} {\bibfield  {journal} {\bibinfo  {journal}
  {Soft Matter}\ }\textbf {\bibinfo {volume} {14}},\ \bibinfo {pages} {9368}
  (\bibinfo {year} {2018})}\BibitemShut {NoStop}%
\bibitem [{\citenamefont {Meyer}\ \emph {et~al.}(2017)\citenamefont {Meyer},
  \citenamefont {Voigtmann},\ and\ \citenamefont {Schilling}}]{Schilling_2017}%
  \BibitemOpen
  \bibfield  {author} {\bibinfo {author} {\bibfnamefont {H.}~\bibnamefont
  {Meyer}}, \bibinfo {author} {\bibfnamefont {T.}~\bibnamefont {Voigtmann}}, \
  and\ \bibinfo {author} {\bibfnamefont {T.}~\bibnamefont {Schilling}},\ }\href
  {\doibase 10.1063/1.5006980} {\bibfield  {journal} {\bibinfo  {journal} {The
  Journal of Chemical Physics}\ }\textbf {\bibinfo {volume} {147}},\ \bibinfo
  {pages} {214110} (\bibinfo {year} {2017})},\ \Eprint
  {http://arxiv.org/abs/https://doi.org/10.1063/1.5006980}
  {https://doi.org/10.1063/1.5006980} \BibitemShut {NoStop}%
\bibitem [{\citenamefont {Meyer}\ \emph {et~al.}(2019)\citenamefont {Meyer},
  \citenamefont {Voigtmann},\ and\ \citenamefont {Schilling}}]{Schilling_2019}%
  \BibitemOpen
  \bibfield  {author} {\bibinfo {author} {\bibfnamefont {H.}~\bibnamefont
  {Meyer}}, \bibinfo {author} {\bibfnamefont {T.}~\bibnamefont {Voigtmann}}, \
  and\ \bibinfo {author} {\bibfnamefont {T.}~\bibnamefont {Schilling}},\ }\href
  {\doibase 10.1063/1.5090450} {\bibfield  {journal} {\bibinfo  {journal} {The
  Journal of Chemical Physics}\ }\textbf {\bibinfo {volume} {150}},\ \bibinfo
  {pages} {174118} (\bibinfo {year} {2019})},\ \Eprint
  {http://arxiv.org/abs/https://doi.org/10.1063/1.5090450}
  {https://doi.org/10.1063/1.5090450} \BibitemShut {NoStop}%
\bibitem [{\citenamefont {Rey}\ and\ \citenamefont
  {Guardia}(1992)}]{rey_dynamical_1992}%
  \BibitemOpen
  \bibfield  {author} {\bibinfo {author} {\bibfnamefont {R.}~\bibnamefont
  {Rey}}\ and\ \bibinfo {author} {\bibfnamefont {E.}~\bibnamefont {Guardia}},\
  }\href {\doibase 10.1021/j100190a104} {\bibfield  {journal} {\bibinfo
  {journal} {The Journal of Physical Chemistry}\ }\textbf {\bibinfo {volume}
  {96}},\ \bibinfo {pages} {4712} (\bibinfo {year} {1992})}\BibitemShut
  {NoStop}%
\bibitem [{\citenamefont {Mullen}\ \emph {et~al.}(2014)\citenamefont {Mullen},
  \citenamefont {Shea},\ and\ \citenamefont
  {Peters}}]{mullen_transmission_2014}%
  \BibitemOpen
  \bibfield  {author} {\bibinfo {author} {\bibfnamefont {R.~G.}\ \bibnamefont
  {Mullen}}, \bibinfo {author} {\bibfnamefont {J.-E.}\ \bibnamefont {Shea}}, \
  and\ \bibinfo {author} {\bibfnamefont {B.}~\bibnamefont {Peters}},\ }\href
  {\doibase 10.1021/ct4009798} {\bibfield  {journal} {\bibinfo  {journal}
  {Journal of Chemical Theory and Computation}\ }\textbf {\bibinfo {volume}
  {10}},\ \bibinfo {pages} {659} (\bibinfo {year} {2014})}\BibitemShut
  {NoStop}%
\bibitem [{\citenamefont {Rosenberg}\ \emph {et~al.}(1980)\citenamefont
  {Rosenberg}, \citenamefont {Berne},\ and\ \citenamefont
  {Chandler}}]{rosenberg1980}%
  \BibitemOpen
  \bibfield  {author} {\bibinfo {author} {\bibfnamefont {R.~O.}\ \bibnamefont
  {Rosenberg}}, \bibinfo {author} {\bibfnamefont {B.~J.}\ \bibnamefont
  {Berne}}, \ and\ \bibinfo {author} {\bibfnamefont {D.}~\bibnamefont
  {Chandler}},\ }\href@noop {} {\bibfield  {journal} {\bibinfo  {journal}
  {Chemical Physics Letters}\ }\textbf {\bibinfo {volume} {75}},\ \bibinfo
  {pages} {162} (\bibinfo {year} {1980})}\BibitemShut {NoStop}%
\bibitem [{\citenamefont {de~Sancho}\ \emph {et~al.}(2014)\citenamefont
  {de~Sancho}, \citenamefont {Sirur},\ and\ \citenamefont
  {Best}}]{de_sancho_molecular_2014}%
  \BibitemOpen
  \bibfield  {author} {\bibinfo {author} {\bibfnamefont {D.}~\bibnamefont
  {de~Sancho}}, \bibinfo {author} {\bibfnamefont {A.}~\bibnamefont {Sirur}}, \
  and\ \bibinfo {author} {\bibfnamefont {R.~B.}\ \bibnamefont {Best}},\ }\href
  {\doibase 10.1038/ncomms5307} {\bibfield  {journal} {\bibinfo  {journal}
  {Nature Communications}\ }\textbf {\bibinfo {volume} {5}} (\bibinfo {year}
  {2014}),\ 10.1038/ncomms5307}\BibitemShut {NoStop}%
\bibitem [{\citenamefont {Daldrop}\ \emph
  {et~al.}(2018{\natexlab{a}})\citenamefont {Daldrop}, \citenamefont {Kappler},
  \citenamefont {Brünig},\ and\ \citenamefont {Netz}}]{daldrop_butane_2018}%
  \BibitemOpen
  \bibfield  {author} {\bibinfo {author} {\bibfnamefont {J.~O.}\ \bibnamefont
  {Daldrop}}, \bibinfo {author} {\bibfnamefont {J.}~\bibnamefont {Kappler}},
  \bibinfo {author} {\bibfnamefont {F.~N.}\ \bibnamefont {Brünig}}, \ and\
  \bibinfo {author} {\bibfnamefont {R.~R.}\ \bibnamefont {Netz}},\ }\href
  {\doibase 10.1073/pnas.1722327115} {\bibfield  {journal} {\bibinfo  {journal}
  {Proceedings of the National Academy of Sciences}\ }\textbf {\bibinfo
  {volume} {115}},\ \bibinfo {pages} {5169} (\bibinfo {year}
  {2018}{\natexlab{a}})}\BibitemShut {NoStop}%
\bibitem [{\citenamefont {Mason}\ and\ \citenamefont
  {Weitz}(1995)}]{Mason1995}%
  \BibitemOpen
  \bibfield  {author} {\bibinfo {author} {\bibfnamefont {T.~G.}\ \bibnamefont
  {Mason}}\ and\ \bibinfo {author} {\bibfnamefont {D.~A.}\ \bibnamefont
  {Weitz}},\ }\href@noop {} {\bibfield  {journal} {\bibinfo  {journal}
  {Physical Review Letters}\ }\textbf {\bibinfo {volume} {74}},\ \bibinfo
  {pages} {1250} (\bibinfo {year} {1995})}\BibitemShut {NoStop}%
\bibitem [{\citenamefont {Lesnicki}\ \emph {et~al.}(2016)\citenamefont
  {Lesnicki}, \citenamefont {Vuilleumier}, \citenamefont {Carof},\ and\
  \citenamefont {Rotenberg}}]{lesnicki_molecular_2016}%
  \BibitemOpen
  \bibfield  {author} {\bibinfo {author} {\bibfnamefont {D.}~\bibnamefont
  {Lesnicki}}, \bibinfo {author} {\bibfnamefont {R.}~\bibnamefont
  {Vuilleumier}}, \bibinfo {author} {\bibfnamefont {A.}~\bibnamefont {Carof}},
  \ and\ \bibinfo {author} {\bibfnamefont {B.}~\bibnamefont {Rotenberg}},\
  }\href {\doibase 10.1103/PhysRevLett.116.147804} {\bibfield  {journal}
  {\bibinfo  {journal} {Physical Review Letters}\ }\textbf {\bibinfo {volume}
  {116}} (\bibinfo {year} {2016}),\ 10.1103/PhysRevLett.116.147804}\BibitemShut
  {NoStop}%
\bibitem [{\citenamefont {Daldrop}\ \emph {et~al.}(2017)\citenamefont
  {Daldrop}, \citenamefont {Kowalik},\ and\ \citenamefont
  {Netz}}]{daldrop_external_2017}%
  \BibitemOpen
  \bibfield  {author} {\bibinfo {author} {\bibfnamefont {J.~O.}\ \bibnamefont
  {Daldrop}}, \bibinfo {author} {\bibfnamefont {B.~G.}\ \bibnamefont
  {Kowalik}}, \ and\ \bibinfo {author} {\bibfnamefont {R.~R.}\ \bibnamefont
  {Netz}},\ }\href {\doibase 10.1103/PhysRevX.7.041065} {\bibfield  {journal}
  {\bibinfo  {journal} {Physical Review X}\ }\textbf {\bibinfo {volume} {7}}
  (\bibinfo {year} {2017}),\ 10.1103/PhysRevX.7.041065}\BibitemShut {NoStop}%
\bibitem [{\citenamefont {Berner}\ \emph {et~al.}(2018)\citenamefont {Berner},
  \citenamefont {M\"uller}, \citenamefont {Gomez-Solano}, \citenamefont
  {Kr\"uger},\ and\ \citenamefont {Bechinger}}]{Bechinger2018}%
  \BibitemOpen
  \bibfield  {author} {\bibinfo {author} {\bibfnamefont {J.}~\bibnamefont
  {Berner}}, \bibinfo {author} {\bibfnamefont {B.}~\bibnamefont {M\"uller}},
  \bibinfo {author} {\bibfnamefont {J.~R.}\ \bibnamefont {Gomez-Solano}},
  \bibinfo {author} {\bibfnamefont {M.}~\bibnamefont {Kr\"uger}}, \ and\
  \bibinfo {author} {\bibfnamefont {C.}~\bibnamefont {Bechinger}},\ }\href@noop
  {} {\bibfield  {journal} {\bibinfo  {journal} {Nature Communications}\
  }\textbf {\bibinfo {volume} {9}},\ \bibinfo {pages} {999} (\bibinfo {year}
  {2018})}\BibitemShut {NoStop}%
\bibitem [{\citenamefont {Selmeczi}\ \emph {et~al.}(2005)\citenamefont
  {Selmeczi}, \citenamefont {Mosler}, \citenamefont {Hagedorn}, \citenamefont
  {Larsen},\ and\ \citenamefont {Flyvbjerg}}]{selmeczi2005}%
  \BibitemOpen
  \bibfield  {author} {\bibinfo {author} {\bibfnamefont {D.}~\bibnamefont
  {Selmeczi}}, \bibinfo {author} {\bibfnamefont {S.}~\bibnamefont {Mosler}},
  \bibinfo {author} {\bibfnamefont {P.~H.}\ \bibnamefont {Hagedorn}}, \bibinfo
  {author} {\bibfnamefont {N.~B.}\ \bibnamefont {Larsen}}, \ and\ \bibinfo
  {author} {\bibfnamefont {H.}~\bibnamefont {Flyvbjerg}},\ }\href {\doibase
  https://doi.org/10.1529/biophysj.105.061150} {\bibfield  {journal} {\bibinfo
  {journal} {Biophysical Journal}\ }\textbf {\bibinfo {volume} {89}},\ \bibinfo
  {pages} {912 } (\bibinfo {year} {2005})}\BibitemShut {NoStop}%
\bibitem [{\citenamefont {Wilemski}\ and\ \citenamefont
  {Fixman}(1974)}]{wilemski_diffusioncontrolled_1974}%
  \BibitemOpen
  \bibfield  {author} {\bibinfo {author} {\bibfnamefont {G.}~\bibnamefont
  {Wilemski}}\ and\ \bibinfo {author} {\bibfnamefont {M.}~\bibnamefont
  {Fixman}},\ }\href {\doibase 10.1063/1.1681163} {\bibfield  {journal}
  {\bibinfo  {journal} {The Journal of Chemical Physics}\ }\textbf {\bibinfo
  {volume} {60}},\ \bibinfo {pages} {878} (\bibinfo {year} {1974})}\BibitemShut
  {NoStop}%
\bibitem [{\citenamefont {Szabo}\ \emph {et~al.}(1980)\citenamefont {Szabo},
  \citenamefont {Schulten},\ and\ \citenamefont {Schulten}}]{szabo_first_1980}%
  \BibitemOpen
  \bibfield  {author} {\bibinfo {author} {\bibfnamefont {A.}~\bibnamefont
  {Szabo}}, \bibinfo {author} {\bibfnamefont {K.}~\bibnamefont {Schulten}}, \
  and\ \bibinfo {author} {\bibfnamefont {Z.}~\bibnamefont {Schulten}},\ }\href
  {\doibase 10.1063/1.439715} {\bibfield  {journal} {\bibinfo  {journal} {The
  Journal of Chemical Physics}\ }\textbf {\bibinfo {volume} {72}},\ \bibinfo
  {pages} {4350} (\bibinfo {year} {1980})}\BibitemShut {NoStop}%
\bibitem [{\citenamefont {Dua}\ and\ \citenamefont
  {Adhikari}(2011)}]{Dua_2011}%
  \BibitemOpen
  \bibfield  {author} {\bibinfo {author} {\bibfnamefont {A.}~\bibnamefont
  {Dua}}\ and\ \bibinfo {author} {\bibfnamefont {R.}~\bibnamefont {Adhikari}},\
  }\href {\doibase 10.1088/1742-5468/2011/04/p04017} {\bibfield  {journal}
  {\bibinfo  {journal} {Journal of Statistical Mechanics: Theory and
  Experiment}\ }\textbf {\bibinfo {volume} {2011}},\ \bibinfo {pages} {P04017}
  (\bibinfo {year} {2011})}\BibitemShut {NoStop}%
\bibitem [{\citenamefont {Gowdy}\ \emph {et~al.}(2017)\citenamefont {Gowdy},
  \citenamefont {Batchelor}, \citenamefont {Neelov},\ and\ \citenamefont
  {Paci}}]{gowdy_nonexponential_2017}%
  \BibitemOpen
  \bibfield  {author} {\bibinfo {author} {\bibfnamefont {J.}~\bibnamefont
  {Gowdy}}, \bibinfo {author} {\bibfnamefont {M.}~\bibnamefont {Batchelor}},
  \bibinfo {author} {\bibfnamefont {I.}~\bibnamefont {Neelov}}, \ and\ \bibinfo
  {author} {\bibfnamefont {E.}~\bibnamefont {Paci}},\ }\href {\doibase
  10.1021/acs.jpcb.7b07075} {\bibfield  {journal} {\bibinfo  {journal} {The
  Journal of Physical Chemistry B}\ }\textbf {\bibinfo {volume} {121}},\
  \bibinfo {pages} {9518} (\bibinfo {year} {2017})}\BibitemShut {NoStop}%
\bibitem [{\citenamefont {Guérin}\ \emph {et~al.}(2012)\citenamefont
  {Guérin}, \citenamefont {Bénichou},\ and\ \citenamefont
  {Voituriez}}]{guerin_non-markovian_2012}%
  \BibitemOpen
  \bibfield  {author} {\bibinfo {author} {\bibfnamefont {T.}~\bibnamefont
  {Guérin}}, \bibinfo {author} {\bibfnamefont {O.}~\bibnamefont {Bénichou}},
  \ and\ \bibinfo {author} {\bibfnamefont {R.}~\bibnamefont {Voituriez}},\
  }\href {\doibase 10.1038/nchem.1378} {\bibfield  {journal} {\bibinfo
  {journal} {Nature Chemistry}\ }\textbf {\bibinfo {volume} {4}},\ \bibinfo
  {pages} {568} (\bibinfo {year} {2012})}\BibitemShut {NoStop}%
\bibitem [{\citenamefont {Plotkin}\ and\ \citenamefont
  {Wolynes}(1998)}]{plotkin_non-markovian_1998}%
  \BibitemOpen
  \bibfield  {author} {\bibinfo {author} {\bibfnamefont {S.~S.}\ \bibnamefont
  {Plotkin}}\ and\ \bibinfo {author} {\bibfnamefont {P.~G.}\ \bibnamefont
  {Wolynes}},\ }\href {\doibase 10.1103/PhysRevLett.80.5015} {\bibfield
  {journal} {\bibinfo  {journal} {Physical Review Letters}\ }\textbf {\bibinfo
  {volume} {80}},\ \bibinfo {pages} {5015} (\bibinfo {year}
  {1998})}\BibitemShut {NoStop}%
\bibitem [{\citenamefont {Das}\ and\ \citenamefont
  {Makarov}(2018)}]{Makarov2018}%
  \BibitemOpen
  \bibfield  {author} {\bibinfo {author} {\bibfnamefont {A.}~\bibnamefont
  {Das}}\ and\ \bibinfo {author} {\bibfnamefont {D.~E.}\ \bibnamefont
  {Makarov}},\ }\href@noop {} {\bibfield  {journal} {\bibinfo  {journal} {J.
  Phys. Chem. B}\ }\textbf {\bibinfo {volume} {122}},\ \bibinfo {pages} {9049 }
  (\bibinfo {year} {2018})}\BibitemShut {NoStop}%
\bibitem [{\citenamefont {Grote}\ and\ \citenamefont
  {Hynes}(1980)}]{grote_stable_1980}%
  \BibitemOpen
  \bibfield  {author} {\bibinfo {author} {\bibfnamefont {R.~F.}\ \bibnamefont
  {Grote}}\ and\ \bibinfo {author} {\bibfnamefont {J.~T.}\ \bibnamefont
  {Hynes}},\ }\href {\doibase 10.1063/1.440485} {\bibfield  {journal} {\bibinfo
   {journal} {The Journal of Chemical Physics}\ }\textbf {\bibinfo {volume}
  {73}},\ \bibinfo {pages} {2715} (\bibinfo {year} {1980})}\BibitemShut
  {NoStop}%
\bibitem [{\citenamefont {Carmeli}\ and\ \citenamefont
  {Nitzan}(1982)}]{carmeli_non-markoffian_1982}%
  \BibitemOpen
  \bibfield  {author} {\bibinfo {author} {\bibfnamefont {B.}~\bibnamefont
  {Carmeli}}\ and\ \bibinfo {author} {\bibfnamefont {A.}~\bibnamefont
  {Nitzan}},\ }\href {\doibase 10.1103/PhysRevLett.49.423} {\bibfield
  {journal} {\bibinfo  {journal} {Physical Review Letters}\ }\textbf {\bibinfo
  {volume} {49}},\ \bibinfo {pages} {423} (\bibinfo {year} {1982})}\BibitemShut
  {NoStop}%
\bibitem [{\citenamefont {Straub}\ \emph {et~al.}(1986)\citenamefont {Straub},
  \citenamefont {Borkovec},\ and\ \citenamefont
  {Berne}}]{straub_non-markovian_1986}%
  \BibitemOpen
  \bibfield  {author} {\bibinfo {author} {\bibfnamefont {J.~E.}\ \bibnamefont
  {Straub}}, \bibinfo {author} {\bibfnamefont {M.}~\bibnamefont {Borkovec}}, \
  and\ \bibinfo {author} {\bibfnamefont {B.~J.}\ \bibnamefont {Berne}},\ }\href
  {\doibase 10.1063/1.450425} {\bibfield  {journal} {\bibinfo  {journal} {The
  Journal of Chemical Physics}\ }\textbf {\bibinfo {volume} {84}},\ \bibinfo
  {pages} {1788} (\bibinfo {year} {1986})}\BibitemShut {NoStop}%
\bibitem [{\citenamefont {Talkner}\ and\ \citenamefont
  {Braun}(1988)}]{talkner_transition_1988}%
  \BibitemOpen
  \bibfield  {author} {\bibinfo {author} {\bibfnamefont {P.}~\bibnamefont
  {Talkner}}\ and\ \bibinfo {author} {\bibfnamefont {H.-B.}\ \bibnamefont
  {Braun}},\ }\href {\doibase 10.1063/1.454318} {\bibfield  {journal} {\bibinfo
   {journal} {The Journal of Chemical Physics}\ }\textbf {\bibinfo {volume}
  {88}},\ \bibinfo {pages} {7537} (\bibinfo {year} {1988})}\BibitemShut
  {NoStop}%
\bibitem [{\citenamefont {Pollak}\ \emph {et~al.}(1989)\citenamefont {Pollak},
  \citenamefont {Grabert},\ and\ \citenamefont {Hänggi}}]{pollak_theory_1989}%
  \BibitemOpen
  \bibfield  {author} {\bibinfo {author} {\bibfnamefont {E.}~\bibnamefont
  {Pollak}}, \bibinfo {author} {\bibfnamefont {H.}~\bibnamefont {Grabert}}, \
  and\ \bibinfo {author} {\bibfnamefont {P.}~\bibnamefont {Hänggi}},\ }\href
  {\doibase 10.1063/1.456837} {\bibfield  {journal} {\bibinfo  {journal} {The
  Journal of Chemical Physics}\ }\textbf {\bibinfo {volume} {91}},\ \bibinfo
  {pages} {4073} (\bibinfo {year} {1989})}\BibitemShut {NoStop}%
\bibitem [{\citenamefont {Ianconescu}\ and\ \citenamefont
  {Pollak}(2015)}]{ianconescu_study_2015}%
  \BibitemOpen
  \bibfield  {author} {\bibinfo {author} {\bibfnamefont {R.}~\bibnamefont
  {Ianconescu}}\ and\ \bibinfo {author} {\bibfnamefont {E.}~\bibnamefont
  {Pollak}},\ }\href {\doibase 10.1063/1.4929709} {\bibfield  {journal}
  {\bibinfo  {journal} {The Journal of Chemical Physics}\ }\textbf {\bibinfo
  {volume} {143}},\ \bibinfo {pages} {104104} (\bibinfo {year}
  {2015})}\BibitemShut {NoStop}%
\bibitem [{\citenamefont {Tucker}\ \emph {et~al.}(1991)\citenamefont {Tucker},
  \citenamefont {Tuckerman}, \citenamefont {Berne},\ and\ \citenamefont
  {Pollak}}]{tucker_comparison_1991}%
  \BibitemOpen
  \bibfield  {author} {\bibinfo {author} {\bibfnamefont {S.~C.}\ \bibnamefont
  {Tucker}}, \bibinfo {author} {\bibfnamefont {M.~E.}\ \bibnamefont
  {Tuckerman}}, \bibinfo {author} {\bibfnamefont {B.~J.}\ \bibnamefont
  {Berne}}, \ and\ \bibinfo {author} {\bibfnamefont {E.}~\bibnamefont
  {Pollak}},\ }\href {\doibase 10.1063/1.461603} {\bibfield  {journal}
  {\bibinfo  {journal} {The Journal of Chemical Physics}\ }\textbf {\bibinfo
  {volume} {95}},\ \bibinfo {pages} {5809} (\bibinfo {year}
  {1991})}\BibitemShut {NoStop}%
\bibitem [{\citenamefont {Kappler}\ \emph {et~al.}(2018)\citenamefont
  {Kappler}, \citenamefont {Daldrop}, \citenamefont {Brünig}, \citenamefont
  {Boehle},\ and\ \citenamefont {Netz}}]{kappler_memory-induced_2018}%
  \BibitemOpen
  \bibfield  {author} {\bibinfo {author} {\bibfnamefont {J.}~\bibnamefont
  {Kappler}}, \bibinfo {author} {\bibfnamefont {J.~O.}\ \bibnamefont
  {Daldrop}}, \bibinfo {author} {\bibfnamefont {F.~N.}\ \bibnamefont
  {Brünig}}, \bibinfo {author} {\bibfnamefont {M.~D.}\ \bibnamefont {Boehle}},
  \ and\ \bibinfo {author} {\bibfnamefont {R.~R.}\ \bibnamefont {Netz}},\
  }\href {\doibase 10.1063/1.4998239} {\bibfield  {journal} {\bibinfo
  {journal} {The Journal of Chemical Physics}\ }\textbf {\bibinfo {volume}
  {148}},\ \bibinfo {pages} {014903} (\bibinfo {year} {2018})}\BibitemShut
  {NoStop}%
\bibitem [{\citenamefont {Tolokh}\ \emph {et~al.}(2002)\citenamefont {Tolokh},
  \citenamefont {White}, \citenamefont {Goldman},\ and\ \citenamefont
  {Gray}}]{tolokh_prediction_2002}%
  \BibitemOpen
  \bibfield  {author} {\bibinfo {author} {\bibfnamefont {I.~S.}\ \bibnamefont
  {Tolokh}}, \bibinfo {author} {\bibfnamefont {G.~W.~N.}\ \bibnamefont
  {White}}, \bibinfo {author} {\bibfnamefont {S.}~\bibnamefont {Goldman}}, \
  and\ \bibinfo {author} {\bibfnamefont {C.~G.}\ \bibnamefont {Gray}},\ }\href
  {\doibase 10.1080/00268970210124828} {\bibfield  {journal} {\bibinfo
  {journal} {Molecular Physics}\ }\textbf {\bibinfo {volume} {100}},\ \bibinfo
  {pages} {2351} (\bibinfo {year} {2002})}\BibitemShut {NoStop}%
\bibitem [{\citenamefont {Gottwald}\ \emph {et~al.}(2015)\citenamefont
  {Gottwald}, \citenamefont {Karsten}, \citenamefont {Ivanov},\ and\
  \citenamefont {Kühn}}]{gottwald_parametrizing_2015}%
  \BibitemOpen
  \bibfield  {author} {\bibinfo {author} {\bibfnamefont {F.}~\bibnamefont
  {Gottwald}}, \bibinfo {author} {\bibfnamefont {S.}~\bibnamefont {Karsten}},
  \bibinfo {author} {\bibfnamefont {S.~D.}\ \bibnamefont {Ivanov}}, \ and\
  \bibinfo {author} {\bibfnamefont {O.}~\bibnamefont {Kühn}},\ }\href
  {\doibase 10.1063/1.4922941} {\bibfield  {journal} {\bibinfo  {journal} {The
  Journal of Chemical Physics}\ }\textbf {\bibinfo {volume} {142}},\ \bibinfo
  {pages} {244110} (\bibinfo {year} {2015})}\BibitemShut {NoStop}%
\bibitem [{\citenamefont {Northrup}\ and\ \citenamefont
  {Hynes}(1980)}]{northrup_stable_1980}%
  \BibitemOpen
  \bibfield  {author} {\bibinfo {author} {\bibfnamefont {S.~H.}\ \bibnamefont
  {Northrup}}\ and\ \bibinfo {author} {\bibfnamefont {J.~T.}\ \bibnamefont
  {Hynes}},\ }\href {\doibase 10.1063/1.440484} {\bibfield  {journal} {\bibinfo
   {journal} {The Journal of Chemical Physics}\ }\textbf {\bibinfo {volume}
  {73}},\ \bibinfo {pages} {2700} (\bibinfo {year} {1980})}\BibitemShut
  {NoStop}%
\bibitem [{\citenamefont {Zwanzig}(1973)}]{zwanzig_nonlinear_1973}%
  \BibitemOpen
  \bibfield  {author} {\bibinfo {author} {\bibfnamefont {R.}~\bibnamefont
  {Zwanzig}},\ }\href {\doibase 10.1007/BF01008729} {\bibfield  {journal}
  {\bibinfo  {journal} {Journal of Statistical Physics}\ }\textbf {\bibinfo
  {volume} {9}},\ \bibinfo {pages} {215} (\bibinfo {year} {1973})}\BibitemShut
  {NoStop}%
\bibitem [{\citenamefont {Baron}\ \emph {et~al.}(2007)\citenamefont {Baron},
  \citenamefont {Trzesniak}, \citenamefont {de~Vries}, \citenamefont {Elsener},
  \citenamefont {Marrink},\ and\ \citenamefont {van Gunsteren}}]{Marrink2007}%
  \BibitemOpen
  \bibfield  {author} {\bibinfo {author} {\bibfnamefont {R.}~\bibnamefont
  {Baron}}, \bibinfo {author} {\bibfnamefont {D.}~\bibnamefont {Trzesniak}},
  \bibinfo {author} {\bibfnamefont {A.~H.}\ \bibnamefont {de~Vries}}, \bibinfo
  {author} {\bibfnamefont {A.}~\bibnamefont {Elsener}}, \bibinfo {author}
  {\bibfnamefont {S.~J.}\ \bibnamefont {Marrink}}, \ and\ \bibinfo {author}
  {\bibfnamefont {W.~F.}\ \bibnamefont {van Gunsteren}},\ }\href@noop {}
  {\bibfield  {journal} {\bibinfo  {journal} {ChemPhysChem}\ }\textbf {\bibinfo
  {volume} {8}},\ \bibinfo {pages} {452 } (\bibinfo {year} {2007})}\BibitemShut
  {NoStop}%
\bibitem [{\citenamefont {Potoyan}\ \emph {et~al.}(2013)\citenamefont
  {Potoyan}, \citenamefont {Savelyev},\ and\ \citenamefont
  {Papoian}}]{Papoian2013}%
  \BibitemOpen
  \bibfield  {author} {\bibinfo {author} {\bibfnamefont {D.~A.}\ \bibnamefont
  {Potoyan}}, \bibinfo {author} {\bibfnamefont {A.}~\bibnamefont {Savelyev}}, \
  and\ \bibinfo {author} {\bibfnamefont {G.~A.}\ \bibnamefont {Papoian}},\
  }\href@noop {} {\bibfield  {journal} {\bibinfo  {journal} {WIREs Comput Mol
  Sci}\ }\textbf {\bibinfo {volume} {3}},\ \bibinfo {pages} {69 } (\bibinfo
  {year} {2013})}\BibitemShut {NoStop}%
\bibitem [{\citenamefont {Marx}\ \emph {et~al.}(1999)\citenamefont {Marx},
  \citenamefont {Tuckerman}, \citenamefont {Hutter},\ and\ \citenamefont
  {Parrinello}}]{marx1999}%
  \BibitemOpen
  \bibfield  {author} {\bibinfo {author} {\bibfnamefont {D.}~\bibnamefont
  {Marx}}, \bibinfo {author} {\bibfnamefont {M.~E.}\ \bibnamefont {Tuckerman}},
  \bibinfo {author} {\bibfnamefont {J.}~\bibnamefont {Hutter}}, \ and\ \bibinfo
  {author} {\bibfnamefont {M.}~\bibnamefont {Parrinello}},\ }\href {\doibase
  10.1038/17579} {\bibfield  {journal} {\bibinfo  {journal} {Nature}\ }\textbf
  {\bibinfo {volume} {397}},\ \bibinfo {pages} {601} (\bibinfo {year}
  {1999})}\BibitemShut {NoStop}%
\bibitem [{\citenamefont {Daldrop}\ \emph
  {et~al.}(2018{\natexlab{b}})\citenamefont {Daldrop}, \citenamefont {Saita},
  \citenamefont {Heyden}, \citenamefont {Lorenz-Fonfria}, \citenamefont
  {Heberle},\ and\ \citenamefont {Netz}}]{daldrop2018}%
  \BibitemOpen
  \bibfield  {author} {\bibinfo {author} {\bibfnamefont {J.~O.}\ \bibnamefont
  {Daldrop}}, \bibinfo {author} {\bibfnamefont {M.}~\bibnamefont {Saita}},
  \bibinfo {author} {\bibfnamefont {M.}~\bibnamefont {Heyden}}, \bibinfo
  {author} {\bibfnamefont {V.~A.}\ \bibnamefont {Lorenz-Fonfria}}, \bibinfo
  {author} {\bibfnamefont {J.}~\bibnamefont {Heberle}}, \ and\ \bibinfo
  {author} {\bibfnamefont {R.~R.}\ \bibnamefont {Netz}},\ }\href {\doibase
  10.1038/s41467-017-02669-9} {\bibfield  {journal} {\bibinfo  {journal}
  {Nature Communications}\ }\textbf {\bibinfo {volume} {9}} (\bibinfo {year}
  {2018}{\natexlab{b}}),\ 10.1038/s41467-017-02669-9}\BibitemShut {NoStop}%
\bibitem [{\citenamefont {Chung}\ and\ \citenamefont
  {Eaton}(2013)}]{eaton2013}%
  \BibitemOpen
  \bibfield  {author} {\bibinfo {author} {\bibfnamefont {H.~S.}\ \bibnamefont
  {Chung}}\ and\ \bibinfo {author} {\bibfnamefont {W.~A.}\ \bibnamefont
  {Eaton}},\ }\href@noop {} {\bibfield  {journal} {\bibinfo  {journal}
  {Nature}\ }\textbf {\bibinfo {volume} {502}},\ \bibinfo {pages} {685}
  (\bibinfo {year} {2013})}\BibitemShut {NoStop}%
\bibitem [{\citenamefont {Laleman}\ \emph {et~al.}(2017)\citenamefont
  {Laleman}, \citenamefont {Carlon},\ and\ \citenamefont
  {Orland}}]{orland2017}%
  \BibitemOpen
  \bibfield  {author} {\bibinfo {author} {\bibfnamefont {M.}~\bibnamefont
  {Laleman}}, \bibinfo {author} {\bibfnamefont {E.}~\bibnamefont {Carlon}}, \
  and\ \bibinfo {author} {\bibfnamefont {H.}~\bibnamefont {Orland}},\
  }\href@noop {} {\bibfield  {journal} {\bibinfo  {journal} {J. Chem. Phys.}\
  }\textbf {\bibinfo {volume} {147}},\ \bibinfo {pages} {214103} (\bibinfo
  {year} {2017})}\BibitemShut {NoStop}%
\bibitem [{\citenamefont {Mel’nikov}\ and\ \citenamefont
  {Meshkov}(1986)}]{melnikov_theory_1986}%
  \BibitemOpen
  \bibfield  {author} {\bibinfo {author} {\bibfnamefont {V.~I.}\ \bibnamefont
  {Mel’nikov}}\ and\ \bibinfo {author} {\bibfnamefont {S.~V.}\ \bibnamefont
  {Meshkov}},\ }\href {\doibase 10.1063/1.451844} {\bibfield  {journal}
  {\bibinfo  {journal} {The Journal of Chemical Physics}\ }\textbf {\bibinfo
  {volume} {85}},\ \bibinfo {pages} {1018} (\bibinfo {year}
  {1986})}\BibitemShut {NoStop}%
\bibitem [{\citenamefont {H\"anggi}\ and\ \citenamefont
  {Weiss}(1984)}]{weiss_1984}%
  \BibitemOpen
  \bibfield  {author} {\bibinfo {author} {\bibfnamefont {P.}~\bibnamefont
  {H\"anggi}}\ and\ \bibinfo {author} {\bibfnamefont {U.}~\bibnamefont
  {Weiss}},\ }\href {\doibase 10.1103/PhysRevA.29.2265} {\bibfield  {journal}
  {\bibinfo  {journal} {Phys. Rev. A}\ }\textbf {\bibinfo {volume} {29}},\
  \bibinfo {pages} {2265} (\bibinfo {year} {1984})}\BibitemShut {NoStop}%
\bibitem [{\citenamefont {Carmeli}\ and\ \citenamefont
  {Nitzan}(1984)}]{nitzan_1984}%
  \BibitemOpen
  \bibfield  {author} {\bibinfo {author} {\bibfnamefont {B.}~\bibnamefont
  {Carmeli}}\ and\ \bibinfo {author} {\bibfnamefont {A.}~\bibnamefont
  {Nitzan}},\ }\href {\doibase 10.1103/PhysRevA.29.1481} {\bibfield  {journal}
  {\bibinfo  {journal} {Phys. Rev. A}\ }\textbf {\bibinfo {volume} {29}},\
  \bibinfo {pages} {1481} (\bibinfo {year} {1984})}\BibitemShut {NoStop}%
\bibitem [{\citenamefont {Straub}\ \emph {et~al.}(1985)\citenamefont {Straub},
  \citenamefont {Borkovec},\ and\ \citenamefont {Berne}}]{berne_1985}%
  \BibitemOpen
  \bibfield  {author} {\bibinfo {author} {\bibfnamefont {J.~E.}\ \bibnamefont
  {Straub}}, \bibinfo {author} {\bibfnamefont {M.}~\bibnamefont {Borkovec}}, \
  and\ \bibinfo {author} {\bibfnamefont {B.~J.}\ \bibnamefont {Berne}},\ }\href
  {\doibase 10.1063/1.449172} {\bibfield  {journal} {\bibinfo  {journal} {The
  Journal of Chemical Physics}\ }\textbf {\bibinfo {volume} {83}},\ \bibinfo
  {pages} {3172} (\bibinfo {year} {1985})},\ \Eprint
  {http://arxiv.org/abs/https://doi.org/10.1063/1.449172}
  {https://doi.org/10.1063/1.449172} \BibitemShut {NoStop}%
\bibitem [{\citenamefont {Grabert}\ \emph {et~al.}(1988)\citenamefont
  {Grabert}, \citenamefont {Schramm},\ and\ \citenamefont
  {Ingold}}]{grabert_1988}%
  \BibitemOpen
  \bibfield  {author} {\bibinfo {author} {\bibfnamefont {H.}~\bibnamefont
  {Grabert}}, \bibinfo {author} {\bibfnamefont {P.}~\bibnamefont {Schramm}}, \
  and\ \bibinfo {author} {\bibfnamefont {G.-L.}\ \bibnamefont {Ingold}},\
  }\href {\doibase https://doi.org/10.1016/0370-1573(88)90023-3} {\bibfield
  {journal} {\bibinfo  {journal} {Physics Reports}\ }\textbf {\bibinfo {volume}
  {168}},\ \bibinfo {pages} {115 } (\bibinfo {year} {1988})}\BibitemShut
  {NoStop}%
\bibitem [{\citenamefont {Kappler}\ \emph {et~al.}(2019)\citenamefont
  {Kappler}, \citenamefont {No\'e},\ and\ \citenamefont {Netz}}]{netz_2019}%
  \BibitemOpen
  \bibfield  {author} {\bibinfo {author} {\bibfnamefont {J.}~\bibnamefont
  {Kappler}}, \bibinfo {author} {\bibfnamefont {F.}~\bibnamefont {No\'e}}, \
  and\ \bibinfo {author} {\bibfnamefont {R.~R.}\ \bibnamefont {Netz}},\ }\href
  {\doibase 10.1103/PhysRevLett.122.067801} {\bibfield  {journal} {\bibinfo
  {journal} {Phys. Rev. Lett.}\ }\textbf {\bibinfo {volume} {122}},\ \bibinfo
  {pages} {067801} (\bibinfo {year} {2019})}\BibitemShut {NoStop}%
\bibitem [{\citenamefont {Marx}(2006)}]{marx2006}%
  \BibitemOpen
  \bibfield  {author} {\bibinfo {author} {\bibfnamefont {D.}~\bibnamefont
  {Marx}},\ }\href {\doibase 10.1002/cphc.200600128} {\bibfield  {journal}
  {\bibinfo  {journal} {ChemPhysChem}\ }\textbf {\bibinfo {volume} {7}},\
  \bibinfo {pages} {1848} (\bibinfo {year} {2006})}\BibitemShut {NoStop}%
\bibitem [{\citenamefont {Reimann}\ \emph {et~al.}(1999)\citenamefont
  {Reimann}, \citenamefont {Schmid},\ and\ \citenamefont
  {Hänggi}}]{reimann_universal_1999}%
  \BibitemOpen
  \bibfield  {author} {\bibinfo {author} {\bibfnamefont {P.}~\bibnamefont
  {Reimann}}, \bibinfo {author} {\bibfnamefont {G.~J.}\ \bibnamefont {Schmid}},
  \ and\ \bibinfo {author} {\bibfnamefont {P.}~\bibnamefont {Hänggi}},\ }\href
  {\doibase 10.1103/PhysRevE.60.R1} {\bibfield  {journal} {\bibinfo  {journal}
  {Physical Review E}\ }\textbf {\bibinfo {volume} {60}},\ \bibinfo {pages}
  {R1} (\bibinfo {year} {1999})}\BibitemShut {NoStop}%
\end{thebibliography}
\end{document}